\documentclass[journal=jctcce,manuscript=article]{achemso}
\usepackage[version=3]{mhchem} 
\usepackage{mathtools}
\usepackage{graphicx}
\usepackage{dcolumn}
\usepackage{bm}
\usepackage[utf8]{inputenc}
\usepackage[T1]{fontenc}
\usepackage{newtxmath}
\usepackage{newtxtext}
\usepackage{etoolbox}
\usepackage{physics}
\usepackage{amsmath}
\usepackage{comment}
\usepackage{color}
\usepackage{algpseudocode}
\usepackage{algorithm}

\author{Hideaki Takahashi}
\affiliation{DISAFA, University of Torino, Grugliasco I10095, Italy}
\email{hideaki.takahashi@unito.it}
\author{Raffaele Borrelli}
\affiliation{DISAFA, University of Torino, Grugliasco I10095, Italy}
\email{raffaele.borrelli@unito.it}

\title{Discretization of Structured Bosonic Environments at Finite Temperature by Interpolative Decomposition: Theory and Application}

\keywords{Quantum Dynamics, Tensor-Trains}

\begin{document}

\begin{tocentry}
    Insert graphical abstract.
    \begin{center}
    \includegraphics[width=0.9\textwidth]{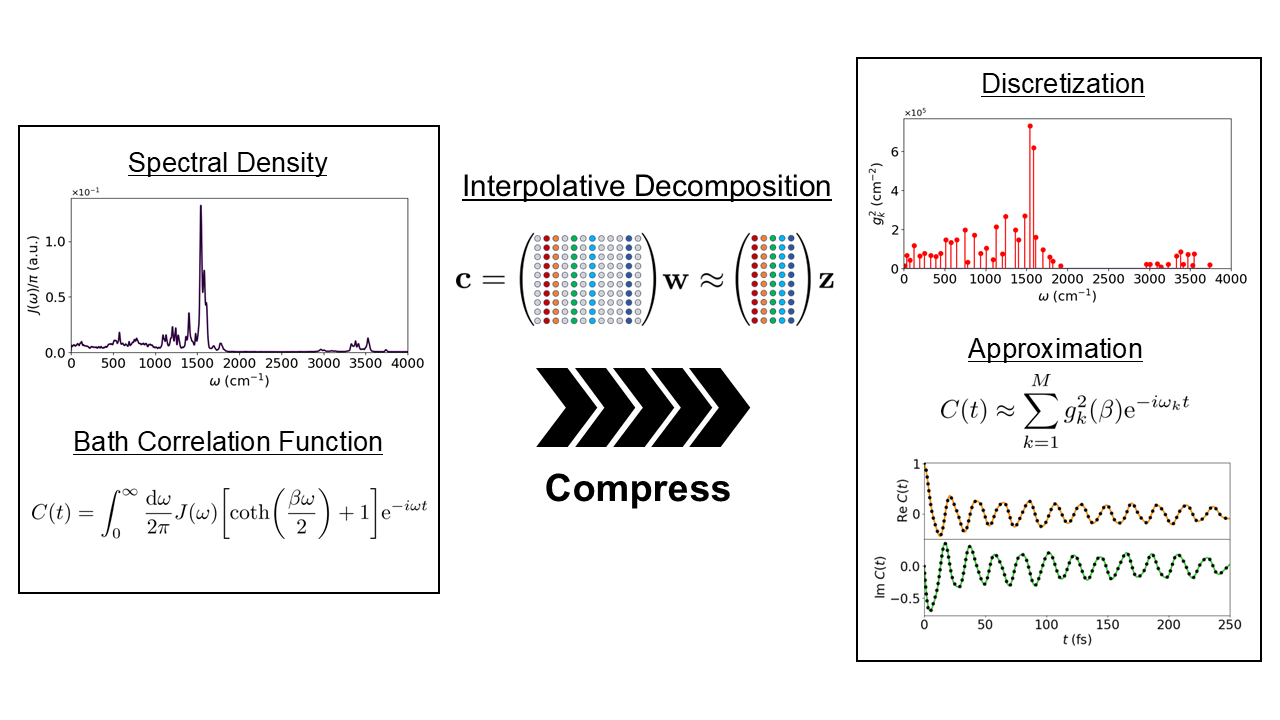}
    \end{center}
\end{tocentry}

\begin{abstract}

We present a comprehensive theory for a novel method to discretize the spectral density of a bosonic heat bath, as introduced in [H. Takahashi and R. Borrelli, J. Chem. Phys. \textbf{161}, 151101 (2024)]. The approach leverages a low-rank decomposition of 
the Fourier-transform relation connecting the bath correlation function to its spectral density.
By capturing the time, frequency, and temperature dependencies encoded in the spectral density-autocorrelation function relation, our method significantly reduces the degrees of freedom required for simulating open quantum system dynamics. 
We benchmark our approach against existing methods and demonstrate its efficacy
through applications to both simple models and a realistic electron transfer process in biological systems. 
Additionally, we show that this new approach can be effectively combined with the tensor-train formalism to investigate the quantum dynamics of systems interacting with complex non-Markovian environments.
Finally, we provide a perspective on the selection and application of various spectral density  discretization techniques.

\end{abstract}

\section{Introduction \label{sec:intro}}
The study of dynamical properties of quantum systems  
interacting with their environments is of fundamental importance across multiple disciplines, including organic electronics,\cite{OberhoferEtAl2017CR,Blumberger2015CR}  
biological systems,\cite{KreisbeckKramer2012JPCL,SchulzeEtAl2016JCP,GelinBorrelli2021JCTC,BorrelliGelin2017SR}  and
quantum information theory.\cite{PalmaEtAl1996PMPES}. Key processes such as exciton and charge transport
in organic materials and decoherence in quantum systems are strongly influenced by environmental degrees of freedom.


Typically, these studies involves coupling the system of interest to bosonic or fermionic environments through a linear coupling and assuming the reservoirs are initially in thermal equilibrium. 
The dynamical properties of these reservoirs are encoded 
in the spectral density (SD) function,
which describe the intensity of the 
noise induced by the bath oscillator at a give frequency, that is the strength of the system-bath coupling.
\cite{KuboStatPhysII1991,ClerkEtAl2010RMP}

Effective approaches for simulating the dynamics of this class of models include, among others, basis set methods—such as the multilayer multiconfiguration time-dependent Hartree (ML-MCTDH)\cite{WangThoss2003JCP} method, time-dependent density matrix renormalization group (TD-DMRG)\cite{RenEtAl2022WCMS} and thermo field dynamics in tensor-train format (TFD-TT)\cite{BorrelliGelin2016JCP, BorrelliGelin2017SR, BorrelliGelin2021WCMS,Borrelli2018CP}—influence-functional based methods—such as the quasi-adiabatic propagator path integral (QUAPI)\cite{MakriMakarov1995JCP,MakriMakarov1995TJoCPa} and hierarchical equations of motion (HEOM)\cite{TanimuraKubo1989JPSJ,Tanimura2020JCP,Borrelli2019JCP}—and chain mapping approaches.\cite{CederbaumEtAl2005PRL,HughesEtAl2009JCP,HughesEtAl2009JCPa, PriorEtAl2010PRL,TamascelliEtAl2019PRL}

Many of these methods often requires the mapping of the continuous SD into a discrete system-bath Hamiltonian.
Several discretization procedures have been proposed for this purpose \cite{HartmannEtAl2019TJoCP}. Equispaced discretization is a very simple method, and has been utilized in several studies \cite{TamuraEtAl2012JCP, SchulzeEtAl2016JCP}, despite its inherent limitations. Logarithmic discretization \cite{BullaEtAl2003PRL, BullaEtAl2008RMP, WangEtAl2016JCP} has been developed to specifically tackle the low-frequency components of Ohmic and sub-Ohmic SDs. Makri\cite{Makri1999JPCB}  has first devised a discretization procedure in which the frequency domain is split into sub-intervals  equally contributing to the reorganization energy. This procedure, which we will refer to as
mode density method (MDM), has been frequently used in the past for the study of model systems\cite{WangEtAl2001JCP} as well as for 
realistic spectral densities.\cite{WaltersEtAl2017JCC}
Alternatively, Gauss quadrature can be used to integrate the bath hybridization function, resulting in a discrete representation of the BCF. In particular, it has been shown that using the SD as a weight function for the quadrature can significantly improve frequency sampling.
This approach is referred to as Bath-Spectral-Density-Orthogonal (BSDO) method.\cite{deVegaEtAl2015PRB, WoodsPlenio2016JMP}

Yet, developing a general method that can provide an optimal number of frequencies and their relative couplings, tailored to a specific temperature, simulation time, and desired accuracy remains an open challenge.  
To tackle this problem, 
Takahashi and Borrelli\cite{TakahashiBorrelli2024} recently introduced a methodology inspired by studies on compressing imaginary-time Green's functions.\cite{OtsukiEtAl2017PRE,ShinaokaEtAl2017PRB,KayeEtAl2022PRB,KayeEtAl2022CPC} This new approach uses the low-rank representation of the Fourier-transform relationship between the SD and the
bath correlation function (BCF), which captures temporal environmental fluctuations, via Interpolative Decomposition (ID) theory.\cite{ChengEtAl2005SJSC,LibertyEtAl2007PNAS,WoolfeEtAl2008AaCHA,HalkoEtAl2011SR}


In this study, we present an in-depth theoretical and computational 
study of the features and performances of this methodology and benchmark it against existing techniques. 
To illustrate its efficacy, we apply it to both simplified model systems and a realistic electron transfer process in a biological system.

The paper is organized as follows.  We present the detailed theory and procedure of the ID approach in Section \ref{sec:theory}.  In Sec. \ref{sec:result1}, the numerical efficiency of the ID approach is demonstrated for the Ohmic and subOhmic spectral densities by comparing it with the other methods, and in Sec. \ref{sec:result2} the discretization of highly structured spectral densities and the dynamics of the electron transfer in cryptochromes are demonstrated.
Conclusions and discussions are reported in Sec. \ref{sec:conclusion}. 

\section{Linear Dissipation in Thermo-field dynamics}
\label{sec:theory}
\subsection{Linear dissipative model}
Let us consider a system linearly coupled to a bosonic heat bath with inverse temperature $\beta$, which consists of a canonical distribution of harmonic oscillators. The Hamiltonian operator can be written as
\begin{equation}
    H = H_\mathrm{S} + \sum_k \omega_{k} a_k^\dagger a_k + V_\mathrm{SB}\sum_{k} g_k  (a_k^\dagger + a_k).
    \label{eq:ham}
\end{equation}
where $H_\mathrm{S}$ is a system Hamiltonian and $V_{\mathrm{SB}}$ is a system part of a system-bath coupling.  $a_{k}, a_{k}^\dagger,\omega_{k}$ and $g_{k}$ are the annihilation operator, creation operator, frequency, and system-bath coupling constant for the $k$th mode of the bath, respectively.  

From a dynamical point of view, the effect of the bosonic heat bath at an inverse temperature $\beta$ on the system is exclusively determined by the BCF\cite{Weiss-book}
\begin{equation}
    C(t)=\frac{1}{2\pi} \int_{-\infty}^{\infty} \dd \omega \;S_\beta(\omega) \mathrm{e}^{-i \omega t}
    \label{eq:fdt1},
\end{equation}
where we define the quantum noise spectral density (QNSD)\cite{ClerkEtAl2010RMP}
\begin{equation}
    S_\beta(\omega)\equiv \frac{1}{2\pi}J(\omega)\qty[\mathrm{coth}\qty(\frac{\beta \omega}{2})+1]
    \label{eq:sbeta}
\end{equation}
with $J(\omega)=\sum_k g_k^2 (\delta(\omega-\omega_k)-\delta(\omega+\omega_k))$.
In the following, we will refer to Eq. \eqref{eq:fdt1} as BCF-QNSD relation.
Explicitly writing 
\begin{align}
    C(t)&=\sum_k \left\{ \frac{g_k^2}{2\pi}\qty[\mathrm{coth}\qty(\frac{\beta \omega_k}{2})+1]\mathrm{e}^{-i\omega_k t} \right. \notag\\
    &\hspace{30pt}\left. - \frac{g_k^2}{2\pi}\qty[-\mathrm{coth}\qty(\frac{\beta \omega_k}{2})+1]\mathrm{e}^{i\omega_k t} \right\},
    \label{eq:bcf-disc}
\end{align}
we can state that when the BCF can be written in the form of Eq. \eqref{eq:bcf-disc}, the corresponding dynamical system is described by the Hamiltonian Eq. \eqref{eq:ham} and the bath is at the inverse temperature $\beta$.

\subsection{Thermo field dynamics}
The latter statement can be transformed into a powerful theoretical and computational tool by exploiting the
Thermo field dynamics (TFD) framework.
TFD is a methodology that facilitates the treatment of quantum systems at non-zero temperatures using the framework of wavefunction formalism\cite{UmezawaEtAl1982,Suzuki1991IJMPB,TakahashiUmezawa1996IJMPB}. Here we only briefly review the TFD approach and demonstrate its relation to the Eq. \eqref{eq:fdt1}.  

We first consider the Hamiltonian of bosonic free particles
\begin{equation}
    H_\mathrm{B} = \sum_{k}\omega_k a_k^\dagger a_k,
\end{equation}  
and label the eigenstates of the $k$-th boson as $\ket{n_k}$.
Then, we introduce the so-called tilde space, denoted as $\ket{\tilde{n}_k}$, which is the Hilbert space of a fictitious dynamical system identical to the original physical system. Additionally, the tensor product of the physical and tilde spaces is referred to as the twin space \cite{Suzuki1991IJMPB,Schmutz1978ZPB}. 
A ket vector in the twin space is given by
\begin{equation}
    \ket{m_k,\tilde{n}_k} \equiv \ket{m_k}\otimes\ket{\tilde{n}_k},
\end{equation}
from which the identity vector is defined as
\begin{equation}
    \ket{I} \equiv \bigotimes_k \sum_{n_k} \ket{n_k,\tilde{n}_k}.
    \label{eq:tfdunit}
\end{equation}
Using the canonical distribution operator $\rho_\mathrm{eq}=\mathrm{e}^{-\beta H_\mathrm{B}}/\mathrm{Tr}\qty{\mathrm{e}^{-\beta H_\mathrm{B}}}$ 
and Eq. \ref{eq:tfdunit}  the so-called thermal vacuum state is derived
\begin{align}
    \label{eq:vac1}
    \ket{0(\beta)} &\equiv \rho_\mathrm{eq}^\frac{1}{2}\ket{I} \\
    &=\prod_k \qty[1-\mathrm{e}^{-\beta\omega_k}]^\frac{1}{2} \exp(\mathrm{e}^{-\beta\omega_k/2} \label{eq:vac2} a_k^\dagger\tilde{a}_k^\dagger)\ket{\mathbf{0}} \\
    &= \mathrm{e}^{-i G_\theta}\ket{\mathbf{0}},\label{eq:vac3}
\end{align}
Here, $\ket{\mathbf{0}}=\bigotimes_{k}\ket{0_k,\tilde{0}_k}$ represents the vacuum state of the ensemble of physical and tilde bosons, and
\begin{equation}
    G_\theta=-i \sum_k \theta_k(\beta)\qty(a_k \tilde{a}_k-a_k^{\dagger} \tilde{a}_k^{\dagger}), \; \theta_k(\beta)=\operatorname{arctanh}\qty(\mathrm{e}^{-\beta \omega_k /2}).
\end{equation}
The transformation $\mathrm{e}^{-i G_\theta}$ is referred to as the thermal Bogoliubov transformation.  It is worth noting that the trasformation from Eq. \eqref{eq:vac2} to Eq. \eqref{eq:vac3} can be performed using the Baker-Campbell-Hausdorff formulas for $\mathrm{su}(1,1)$ Lie algebra.\cite{Ban1993JOSABJ}



Let us now consider the case in which the initial density matrix of our model can be represented as the direct product of a pure state of the system 
$\ketbra{\psi_\mathrm{e}}$,   and of the bath density matrix, that is
\begin{equation}
    \rho(0)=Z^{-1} \ketbra{\psi_\mathrm{e}} \rho_{\mathrm{B}}.
\end{equation}
where $Z$ is a properly defined partition function.
Under this assumption, it is possible to demonstrate \cite{BorrelliGelin2016JCP,BorrelliGelin2017SR,BorrelliGelin2021WCMS,GelinBorrelli2017AdP,deVegaBanuls2015PRA} that
the expectation value of an arbitrary operator $A$ acting on the Hilbert space of the physical system can be written as
\begin{equation}
    \expval{A(t)}=\expval{A_\theta}{\psi_\theta(t)}
\end{equation}
where the wavefunction $\ket{\psi_\theta(t)}$ satisfies the Schrödinger equation
\begin{equation}
    i \pdv{t}\ket{\psi_\theta(t)}=H_\theta\ket{\psi_\theta(t)}, 
    \quad
\ket{\psi_\theta(0)} = \ket{\psi_\mathrm{e}}\otimes\ket{\mathbf{0}}.
    \label{eq:tseq}
\end{equation}
with the thermal operators
\begin{equation}
    H_\theta=\mathrm{e}^{i G} \hat{H} \mathrm{e}^{-i G} \quad A_\theta=\mathrm{e}^{i G} A \mathrm{e}^{-i G} .
  \label{eq:tfdmodham}  
\end{equation}
The modified Hamiltonian operator $\hat{H}$ of Eq. \eqref{eq:tfdmodham} is defined as
\begin{equation}
    \hat{H}=H-\tilde{H}_\mathrm{B}
    \label{eq:barh}
\end{equation}
where $\tilde{H}_{\mathrm{B}} = \sum_k \omega_k \tilde{a}_k^{\dagger} \tilde{a}_k$ is the free-boson Hamiltonian operator of the bosonic tilde space.
The thermal Hamiltonian $H_\theta$ controls the finite temperature dynamics and is readily obtained by applying the Bogoliubov thermal transformation to the Hamiltonian operator of Eq. \eqref{eq:barh}
\begin{align}
    H_\theta= & \mathrm{e}^{iG_\theta} \hat{H} \mathrm{e}^{-iG_\theta} \notag\\
    = & H_\mathrm{S}+\sum_k \omega_k\qty(a_k^{\dagger} a_k-\tilde{a}_k^{\dagger} \tilde{a}_k) \notag\\
    & +V_\mathrm{SB}\sum_{k} g_{k}\qty{\qty(a_{k}+a_{k}^{\dagger}) \cosh \qty(\theta_k)+\qty(\tilde{a}_{k}+\tilde{a}_{k}^{\dagger}) \sinh\qty(\theta_k)}.
    \label{eq:ham-thermal}
\end{align} 

Representing the operators and frequencies of physical space and tilde space with common symbols as
\begin{align}
    \qty{a_k,\tilde{a}_k} &\rightarrow \qty{a_k} \\
    \qty{\omega_k,-\omega_k} & \rightarrow \qty{\omega_k} \\
    \qty{g_k\cosh{\theta_k},g_k\sinh{\theta_k}} & \rightarrow \qty{g_k(\beta)},
\end{align}
the TFD Hamiltonian Eq. \eqref{eq:ham-thermal} can be expressed in the simpler form
\begin{equation}
    H_\theta =  H_\mathrm{S}+\sum_k \omega_k a_k^{\dagger} a_k +V_\mathrm{SB}\sum_{k} g_k(\beta)(a_{k}+a_{k}^{\dagger}).\label{eq:tfdgenericham}
\end{equation}

\subsection{Connection between TFD and the BCF-QNSD relation}
The parameters $g_{k} \cosh (\theta_k)$ and $g_{k} \sinh (\theta_k)$ entering the thermal Hamiltonian $H_\theta$ govern the coupling of the subsystem with the physical and tilde bosonic DoFs. The SD for the thermal system-bath Hamiltonian Eq. \eqref{eq:ham-thermal} can be written as
\begin{equation}
    J_\theta(\omega) = J_p(\omega) + J_t(\omega) 
    \label{eq:sbeta-tfd}
\end{equation}
where 
\begin{align}
    J_p(\omega) &\equiv\sum_k\qty(g_{k} \cosh \qty(\theta_k))^2 \delta\qty(\omega-\omega_k)  \Theta(\omega) \\
    J_t(\omega) &\equiv\sum_k\qty(g_{k} \sinh \qty(\theta_k))^2 \delta\qty(\omega+\omega_k)\Theta(-\omega),
\end{align}
where $\Theta(\omega)$ is the Heaviside step function,
which describe the system-bath couplings in the physical (subscript $p$) and tilde (subscript $t$) subspace, respectively. As the temperature goes to zero, $J_p(\omega) \rightarrow J(\omega)$ and $J_t(\omega) \rightarrow 0$.
By making use of the relations
\begin{align}
    \cosh^2(\theta_k) & = 
    \frac{1}{2}\qty[\coth(\beta\omega_k/2)+1] \\
    \sinh^2(\theta_k) &= 
    -\frac{1}{2}\qty[-\coth(\beta\omega_k/2)+1],
\end{align}
we immediately see that Eq. \eqref{eq:sbeta} and Eq. \eqref{eq:sbeta-tfd} are equivalent as
\begin{equation}
    J_\theta(\omega) = \pi S_\beta(\omega).
\end{equation}
Therefore, we have demonstrated that the system-bath Hamiltonian of Eq. \eqref{eq:ham} at temperature $\beta$, and the extended zero temperature Hamiltonian Eq. \eqref{eq:ham-thermal} have the same correlation function and hence the same dynamical behaviour.

Concluding, we can state that whenever the BCF can be written in the form
\begin{equation}
    C(t) = \sum_k g_k^2(\beta) e^{-i\omega_k t},\;\; \omega_k,g_k(\beta)\in\mathbb{R},
    \label{eq:bcf-disc2}
\end{equation}
the system dynamics at temperature $\beta$ is described by the TFD Hamiltonian of Eq. \eqref{eq:tfdgenericham}.  This is the fundamental principle of our method.  Note that the frequencies $\omega_k$ can be negative in finite temperature cases, and here we do not perform thermo-field doubling, as in the standard TFD formalism\cite{deVegaBanuls2015PRA,BorrelliGelin2016JCP}, but instead directly determine a set of temperature-dependent parameters $(\omega_k, g_k(\beta))$.

\section{Low-rank discretization}
Up to this point, we have discussed two equivalent representations of the linear dissipative model. 
Our goal is finding 
an approximation $\bar C(t)$ of the BCF 
of the form of Eq. \eqref{eq:bcf-disc2} with the least number of summands possible, providing a requested tolerance $\epsilon$ within a specified time interval $[0,T]$.
That is, we wish to solve the problem
\begin{equation}
    \left\|C(t)-\bar C(t)\right\|_\infty < \epsilon,\quad t\in[0,T] :
\quad \tilde{C}(t)=\sum_{k=1}^M g_k^2 \mathrm{e}^{-i\omega_k t}
\label{eq:BCFAPPROX}
\end{equation}
with $g_k, \omega_k \in \mathbb{R}$.
Rather than computing $C(t)$ and directly solve the above minimization problem we 
exploit the information contained in the  Eq. \eqref{eq:fdt1}, 
and reduce its rank by using the Interpolative Decomposition theory.
In the following, we present the fundamental theoretical framework and the detailed procedure of our methodology.



\subsection{Calculating the frequencies $\omega_k$}
\subsubsection{Initial discretization of $f(t,\omega)$}
We begin by representing Eq. \eqref{eq:fdt1} in a form that is suitable for numerical analysis.
First, we set a cutoff frequency $\Omega$ 
such that $S_\beta(\omega) = 0$ outside the interval $[-\Omega,\Omega]$
and a cutoff time $T$.  Then rewrite Eq. \eqref{eq:fdt1} as
\begin{equation}    
    C(t)=\int_{-\Omega}^{\Omega} \dd \omega \;f(t,\omega),\quad t\in [0,T].
    \label{eq:cred}
\end{equation}
where $f(t,\omega)\equiv S_\beta(\omega) \mathrm{e}^{-i \omega t}$.
We now define the sets  $\mathbf{T}=\qty{(t_i)_{i=1}^{m}}$ 
and $\boldsymbol{\Omega}=\qty{(\omega_j)_{j=1}^{n}}$  
where $t_i \in [0,T]$ and $\omega_i \in [-\Omega,\Omega]$,
and create a dense rectangular grid $\mathbf{T}\times\boldsymbol{\Omega}$ on which $f(t,\omega)$ is discretized.
In this study, we employ an equispaced fine grid, with $t_i=(i-1)T/(m-1)$ and $\omega_j=2(j-1)\Omega/(n-1)-\Omega$ with $m,n$ being two large integers.
Then the integral in Eq. \eqref{eq:cred} can be approximated for each $t_i$ using a quadrature
\begin{equation}
    C(t_i) = \sum_{j=1}^n f(t_i,\omega_j)w_j + e_i\label{eq:BCFDISCW}
\end{equation}
where $w_j$ are the weights of a quadrature and $e_i$ are approximation errors.

Using Eq. \eqref{eq:BCFDISCW}
we use the information contained in the 
matrix $f(t_i,\omega_j)$ to obtain the subset of 
frequencies $\omega_k$ defining our discretized model,
and the actual values $C(t_i)$ to determine the system-bath coupling strength. The procedure for obtaining the frequency subset is based on the ID theory.

\begin{figure}[t]
    \centering
    \includegraphics[scale=0.07]{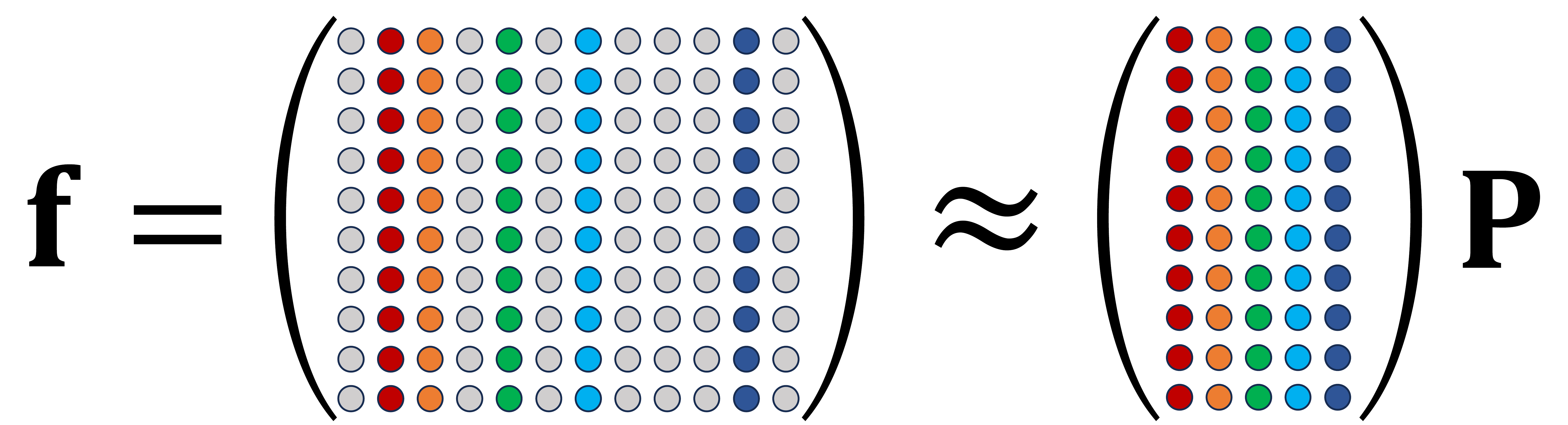}
    \caption{Schematic depiction of the ID Eq. \eqref{eq:id}}
    \label{fig:id}
\end{figure}

\subsubsection{Interpolative Decomposition}
ID is a low-rank approximation method for compressing a matrix.  It is related to other matrix factorization techniques such as singular value decomposition (SVD) and QR decomposition\cite{ChengEtAl2005SJSC,LibertyEtAl2007PNAS,WoolfeEtAl2008AaCHA,HalkoEtAl2011SR}.  One advantage of ID is that it selects and uses a subset of the columns of an original matrix for approximation, which is well-suited to our purpose.
The essential idea of ID is to express a given matrix $\mathbf{A}$ as the product of two matrices.  One is a subset of columns from the original matrix, and the other is a matrix that interpolates the remaining columns. This subset can be chosen to capture the most significant features of $\mathbf{A}$.
Formally, for a given matrix $\mathbf{A}$ of size $m \times n$, interpolative decomposition seeks to find a decomposition such that
\begin{equation}
    \mathbf{A} \approx \mathbf{B}\mathbf{P}
\end{equation}
where $\mathbf{B}$ is an $m \times r$ matrix consisting of $r$ columns selected from $\mathbf{A}$, where $r \leq n$. These columns are chosen so that they provide a good basis for approximating the entire matrix.  $\mathbf{P}$ is a $r \times n$ matrix that contains coefficients such that it approximates $\mathbf{A}$ when multiplied by $\mathbf{B}$. The columns of $\mathbf{P}$ contain a mixture of 1s and 0s that effectively select the corresponding columns in $\mathbf{B}$, and the other entries in $\mathbf{P}$ are coefficients that interpolate the remaining columns of $\mathbf{A}$.

In order to apply ID to our problem, we 
 construct a real matrix $\mathbf{f}_\mathrm{2m\times n}$ with entries $\Re f(t_i,\omega_j)$ and $\Im f(t_i,\omega_j)$
\begin{equation}
    \mathbf{f}_{2m\times n} = \mqty( \qty(\Re f(t_i,\omega_j))_{i,j=1}^{m,n} \\ \qty(\Im f(t_i,\omega_j))_{i,j=1}^{m,n} ) \in\mathbb{R}^{2m\times n}
    \label{eq:matf}
\end{equation}
Then we can construct a rank $r$ ID of $\mathbf{f}_{2m\times n}$ as
\begin{equation}
    \mathbf{f}_{2m \times n} = \mathbf{B}_{2m\times r}\mathbf{P}_{r\times n} + \mathbf{E}_{2m\times n}
    \label{eq:id}
\end{equation}
where
\begin{equation}
    \mathbf{B}_{2m\times r}=\mqty( \qty(\Re f(t_i,\omega_k))_{i,k=1}^{m,r} \vspace{5pt}
    \\ \qty(\Im f(t_i,\omega_k))_{i,k=1}^{m,r} ) \in\mathbb{R}^{2m\times r}
\end{equation}
with $\omega_k \in \tilde{\boldsymbol{\Omega}} \subset \boldsymbol{\Omega}$, 
$\mathbf{P}_{r\times n}\in\mathbb{R}^{r\times n}$, and $\mathbf{E}_{2m \times n}\in\mathbb{R}^{2m \times n}$ is an error matrix  with $\|\mathbf{E}_{2m \times n}\|\leq\epsilon$.  Eq. \eqref{eq:id} is schematically depicted in FIG. \ref{fig:id}.

The subset $\tilde{\boldsymbol{\Omega}}$ corresponds to the selected columns in the ID.  Entrywise, Eq. \eqref{eq:id} can be written as
\begin{align}
    f(t_i,\omega_j)&=\sum_{k=1}^r f(t_i,\omega_k)P_{kj}+E_{ij}' \label{eq:fapprox1}
\end{align}
where $P_{kj}=(\mathbf{P}_{r\times n})_{kj}$, $\mathrm{Re} E_{ij}'=(\mathbf{E}_{2m\times n})_{ij}$ and $\mathrm{Im} E_{ij}'=(\mathbf{E}_{2m \times n})_{(2i)j}$ are entries of matrices $\mathbf{Z}_{r \times n}$ and $\mathbf{E}_{2m \times n}$.
Substituting Eq. \eqref{eq:fapprox1} into Eq. \eqref{eq:BCFDISCW}, we
obtain
\begin{align}
    C(t_i)&=\sum_{j=1}^m\sum_{k=1}^r f\qty(t_i,\omega_k) P_{kj} w_j + \sum_{j=1}^m E'_{ij}w_j  + e_i \notag \\
    &= \sum_{k=1}^r f\qty(t_i,\omega_k) z_k  + e'_i
    \label{eq:ct}
\end{align}
where $z_k = \sum_j P_{kj}w_j$ and $e'_i=e_i+\sum_{j=1}^m E'_{ij}w_j$.  
\footnote{Note that using a complex matrix for ID, without separating the real and imaginary parts of $f(t,\omega)$, is not feasible.
This occurs because the frequencies must be chosen such that the matrix $P$, which leads to the coefficients $z_k$, is real.}
At this stage of the procedure, a specific subset of $r$ frequencies $\omega_k$ has been selected out of the initial set $\boldsymbol{\Omega}$ to 
represent the discretized bath.
The final step involves determining the
system-bath coupling parameters $g_k(\beta)$, which requires evaluating
the weights, $z_k$.

\subsection{Calculating the coupling strengths $g_k(\beta)$}
Let us now define the vector $\mathbf{c}$ that stores the values of the BCF for each sample time
\begin{equation}
    \mathbf{c}=\mqty( \qty(\Re C(t_{i}))_{i=1}^{m} \vspace{5pt} \\ 
    \qty(\Im C(t_{i}))_{i=1}^{m} )  \in\mathbb{R}^{2m},
    \label{eq:vecc}
\end{equation}
then, from Eq. \eqref{eq:ct}, the coefficients $\mathbf{z}=(z_k)_{k=1}^{r}\in \mathbb{R}_{+}^{r}$ can be obtained by solving the overdetermined linear system 
\begin{equation}
    \min_\mathbf{z}\left\|\mathbf{c}-\mathbf{B}\mathbf{z}\right\|_2^2 \quad\text{s.t.} \quad z_k \geq 0.
\end{equation}
This problem can be solved using the so-called nonnegative least squares (NNLS) technique\cite{Lawson1974}. 
Finally, we obtain the desired approximation
 \begin{align}
     \bar{C}(t)&=\sum_{k=1}^{M} f(t,\omega_k)z_k,\quad z_k > 0\notag\\
     &=\sum_{k=1}^{M}z_k S_\beta(\omega_k) \mathrm{e}^{-i \omega_k t},\quad z_k > 0.
     \label{eq:cedr}
 \end{align}
 where, for the Hamiltonian operator to be Hermitian, the weights $z_k$ must be positive.  From the approximated BCF Eq. \eqref{eq:cedr}, we can construct a temperature-dependent effective Hamiltonian Eq. \eqref{eq:tfdgenericham} by setting 
 \begin{equation}
     g_k(\beta)=\sqrt{z_k S_\beta(\omega_k)}.
 \end{equation} 
Since the NNLS procedure may result in some $z_k$ to be zero, in the general case $M\leq r$.  

The overall accuracy of the method is determined by the accuracy of the ID and of the NNLS.
We note that ID induces a pivoting of the columns of $\mathbf{f}$; therefore, the $M$ frequencies $\qty{\omega_k}$ do not correspond to the first $M$ frequencies of the initial discretization grid, but rather form a special subset of it.  The discrete data of $C(t)$ in Eq. \eqref{eq:vecc} for the NNLS procedure can be obtained through numerical integration, which can be combined with interpolation techniques such as spline interpolation or the AAA algorithm.\cite{NakatsukasaEtAl2018SJSC}

\subsection{Actual usage of the ID method}
The method discussed above can be applied
in two different ways: i) setting the threshold $\epsilon$ of ID ii) setting the maximum rank of the matrix $B$ in Eq \eqref{eq:id}.  The parameters required as an input are (1) the cutoff frequency $\Omega$, (2) the cutoff time $T$, (3) the number of points used in the initial discretization, (4) the threshold $\epsilon$ or the rank $r$.

\begin{figure}[t]
    \centering
    \includegraphics[width=8.4cm]{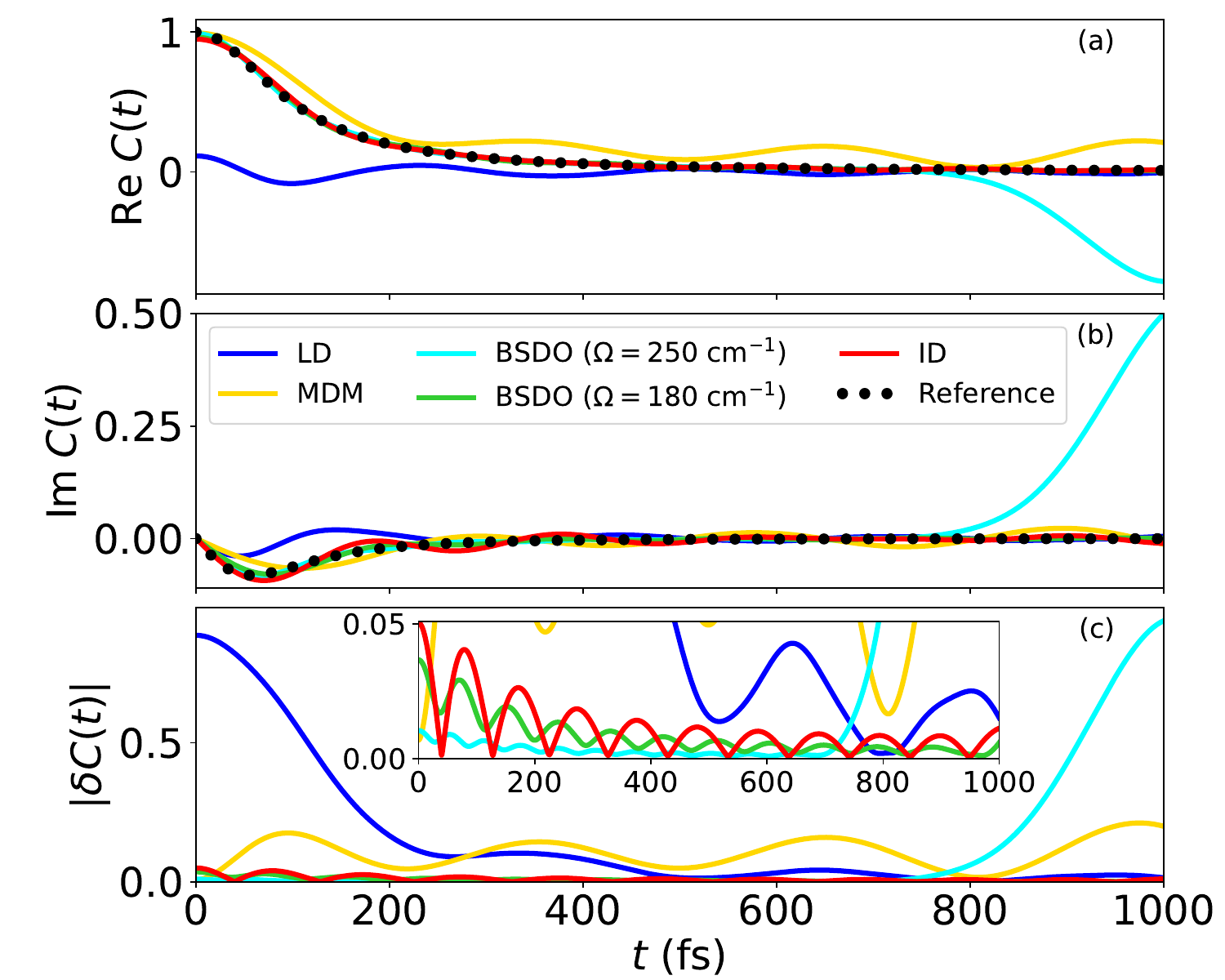}
    \caption{(a) Real part, (b) imaginary part and (c) absolute error of the bath correlation function for the Ohmic spectral density approximated using the LD (blue), MDM (yellow) and ID (red), BSDO for the frequency interval $[-250,250]\;\mathrm{cm}^{-1}$ (cyan) and $[-180,180]\;\mathrm{cm}^{-1}$ (green).  All the BCFs are approximated with $20$ sample points.  The absolute error is defined as $|\delta C(t)|=|C_\mathrm{approx}(t)-C(t)|$.  Black dotted lines are the references.  All the results are normalized to the absolute value of $C(0)$.}
    \label{fig:bcf-ohmic}
\end{figure}

\begin{figure}[t]
    \centering
    \includegraphics[width=8.4cm]{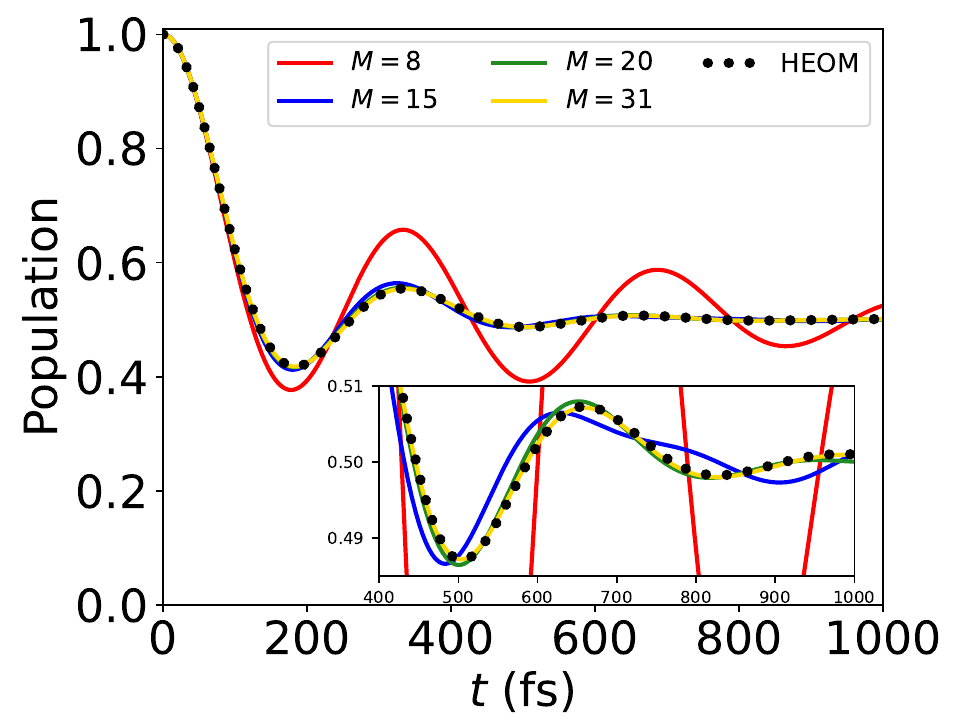}
    \caption{Population dynamics of the first state of the spin-boson model Eq. \eqref{eq:spin-boson} with the Ohmic spectral density calculated using 8 (red), 15 (blue), 20 (green) and 31 (yellow) bosons obtained using the ID approach.  The dotted line is the result calculated using HEOM with the ESPRIT decomposition.}
    \label{fig:pop-ohmic1}
\end{figure}

\section{Discretization of model spectral densities \label{sec:result1}}
As a simple example, we first consider spectral densities in power law form with an exponential cut-off
\begin{equation}
    \label{eq:sd_pl}
    J(\omega) = \pi\alpha\omega_c^{1-s}\omega^s\mathrm{e}^{-\omega/\omega_c},
\end{equation}
which has been extensively employed to study a variety of open quantum systems\cite{LeggettEtAl1987RMP}.  A heat bath is classified according to the value of the parameter $s$ as Ohmic ($s=1$), sub-Ohmic ($0<s<1$), and super-Ohmic ($s>1$). In order to assess the performance of our method, we compare it with  LD\cite{BullaEtAl2003PRL,BullaEtAl2008RMP,WangEtAl2016JCP}, MDM\cite{Makri1999JPCB,WangEtAl2001JCP,WaltersEtAl2017JCC} and BSDO\cite{deVegaEtAl2015PRB,WoodsPlenio2016JMP} approaches (See Appendix for details).  Note that the LD and BSDO approaches are applied to the QNSD, although they were originally developed for the SD, to allow a fair comparison with the ID approach.  It should be noted that the BSDO approach for QNSD is equivalent to the thermalized version of the time-evolving density operator with orthogonal polynomials (T-TEDOPA)\cite{TamascelliEtAl2019PRL} through chain mapping. 

For testing population dynamics, we adopt the spin-boson model
\begin{equation}
    H = -\Delta \sigma_x + \sum_k \omega_k a_k^\dagger a_k + \sigma_z\sum_kg_k\qty(a_k^\dagger + a_k)
    \label{eq:spin-boson}
\end{equation}
where $\sigma_x$ and $\sigma_z$ are the Pauli matrices, $\Delta$ is the coupling between states. 
The parameters are fixed to $\omega_c=53\;\mathrm{cm}^{-1}$, $\alpha=5$ and $\Delta=40\;\mathrm{cm}^{-1}$.  Note that the parameters used in this section are similar to those employed in our previous study on the decomposition of BCFs.\cite{TakahashiEtAl2024TJoCPa}
The initial grids for the ID approach comprise 500 points in time domain and 2000 points in frequency domain for all the results in this section.  The results obtained using the HEOM approach with the ESPRIT technique for the decomposition of the BCF\cite{TakahashiEtAl2024TJoCPa,TakahashiEtAl2024TJoCP,
RoyKailath1989ITASSPa,PichotEtAl2013IJMS} are also shown in the figures as references.

\subsection{Ohmic bath}
We present the BCF and population dynamics for the Ohmic case ($s=1$) in FIG. \ref{fig:bcf-ohmic} and \ref{fig:pop-ohmic1}, respectively.   The temperature is set to $300\;\text{K}$.  The cutoff time and cutoff frequency for the ID approach are $T = 1000\;\text{fs}$ and $\Omega = 500\;\mathrm{cm}^{-1}$, respectively.  The BCF is approximated using 20 sample points, which allow the population dynamics to converge with the ID approach (see FIG. \ref{fig:pop-ohmic1}).  We also present the results obtained using the LD, MDM, and BSDO approaches. For the BSDO method, we provide two results using different frequency intervals. The result obtained for the frequency interval $[-250, 250]\; \mathrm{cm}^{-1}$ is the most accurate at early times, but around $t=700\; \mathrm{fs}$, the error starts to increase. In contrast, the interval $[-180, 180]\; \mathrm{cm}^{-1}$ yields less accurate results initially but performs better over the whole time interval. Thus, the performance of the BSDO approach is very sensitive to the choice of the frequency interval, which must be carefully defined. Conversely, the ID approach can automatically determine the optimal frequencies to represent the BCF with a specified accuracy, as long as a sufficiently wide frequency interval is used.
If the interval is properly chosen, the ID and BSDO approaches show good performance in approximating the BCF over the entire time range.  On the other hand, the LD and MDM approaches return very large errors with 20 sample points and require more than 100 points to achieve comparable accuracy.  See the SM for the details.  In FIG. \ref{fig:pop-ohmic1}, we observe that the dynamics rapidly converge with increasing number of sample points.  The BCF calculated using the ID approach with $20$ vibrations is already in good agreement with the HEOM result.  In contrast, in a previous study \cite{BorrelliGelin2016JCP}, the same dynamics has been calculated using the MD method with 200 sample points.

\subsection{Sub-Ohmic bath}
\begin{figure}[t]
    \centering
    \includegraphics[width=8.4cm]{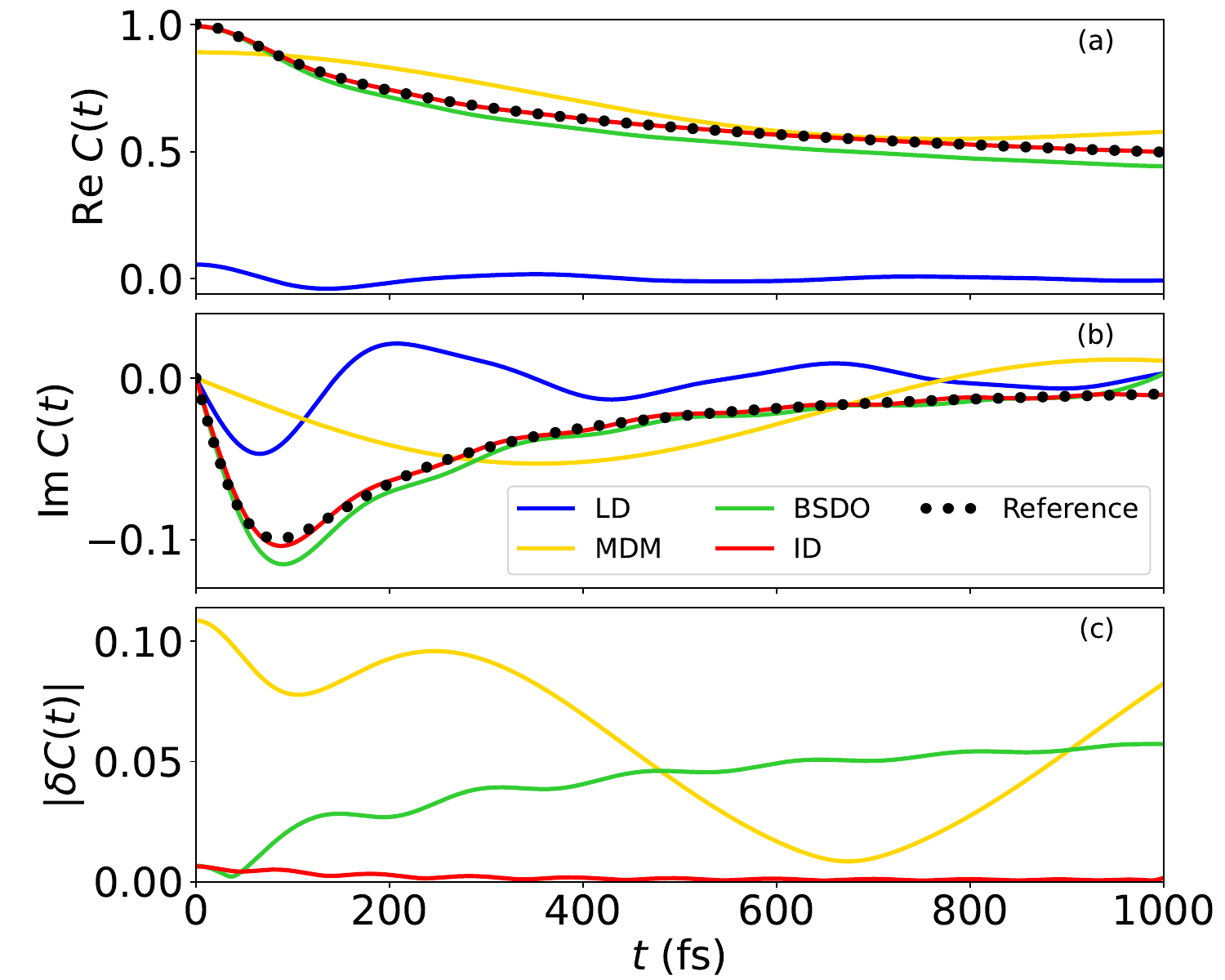}
    \caption{(a) Real part, (b) imaginary part and (c) absolute error of the bath correlation function for the sub-Ohmic spectral density approximated using the LD (blue), MDM (yellow), BSDO (green) and ID (red) with $20$ sample points.  The absolute error is defined as $|\delta C(t)|=|C_\mathrm{approx}(t)-C(t)|$.  Black dotted lines are the references.  All the results are normalized to the absolute value of $C(0)$.}
    \label{fig:bcf-subohmic}
\end{figure}

\begin{figure}[t]
    \centering
    \includegraphics[width=8.4cm]{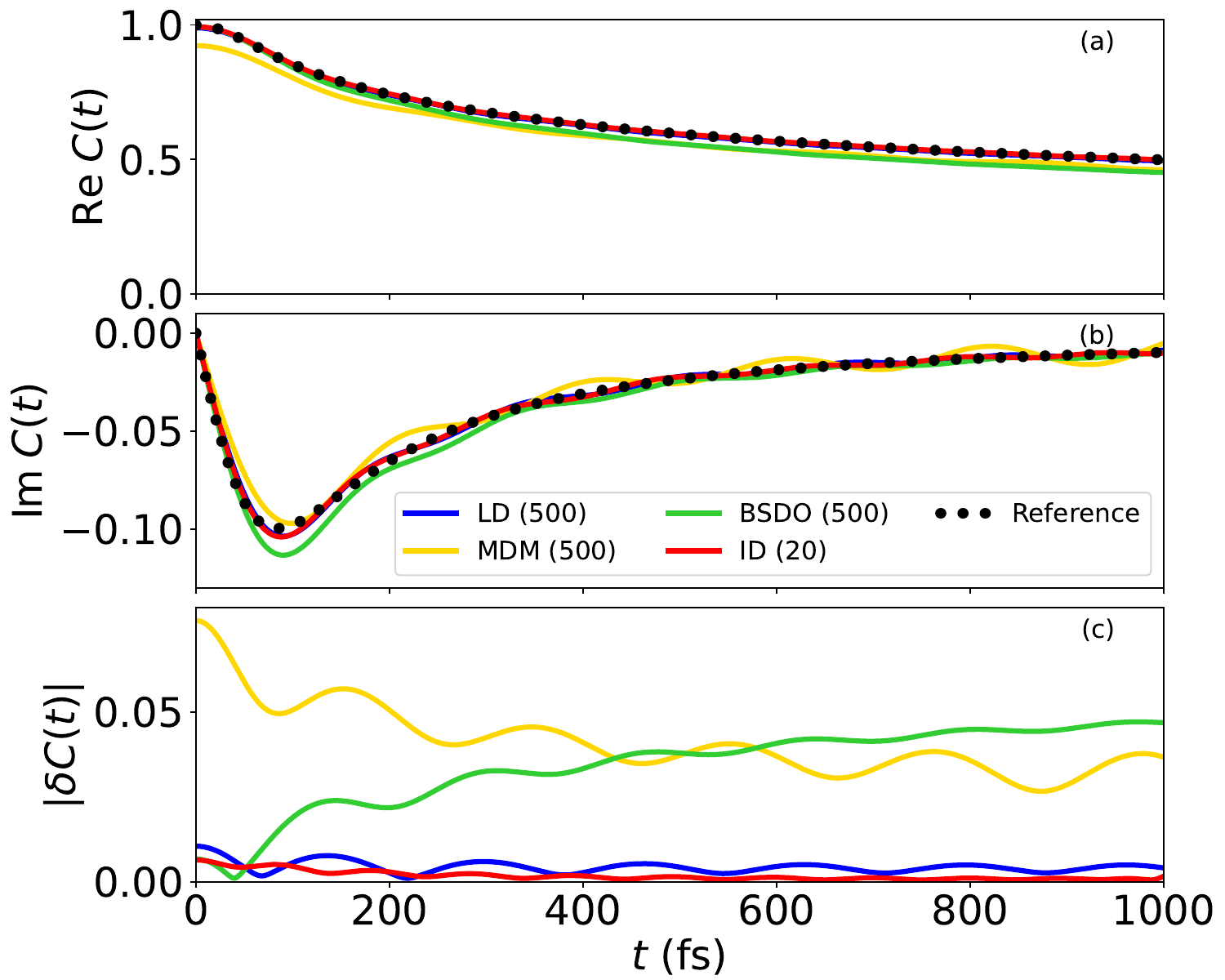}
    \caption{(a) Real part, (b) imaginary part and (c) absolute error of the bath correlation function for the sub-Ohmic spectral density approximated using the LD (blue), MDM (yellow), BSDO (green) and ID (red) with $20$ sample points.  The absolute error is defined as $|\delta C(t)|=|C_\mathrm{approx}(t)-C(t)|$.  Black dotted lines are the references.  All the results are normalized to the absolute value of $C(0)$.}
    \label{fig:bcf-subohmic2}
\end{figure}

\begin{figure}[t]
    \centering
    \includegraphics[width=8.4cm]{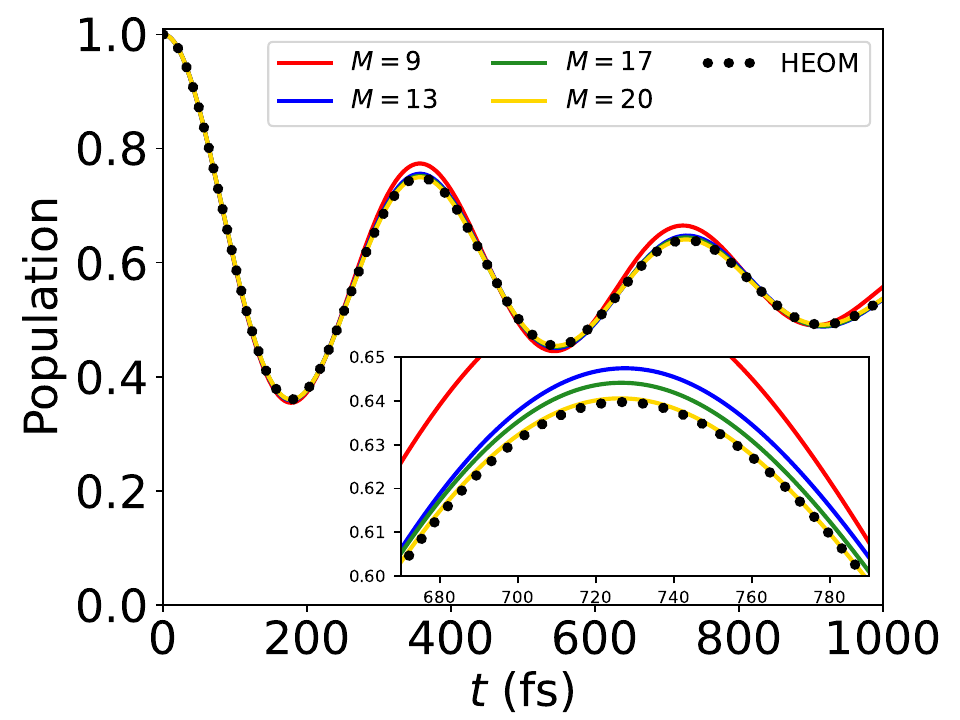}
    \caption{Population dynamics of the first state of the spin-boson model Eq. \eqref{eq:spin-boson} with the sub-Ohmic spectral density calculated with 9 (red), 13 (blue), 17 (green) and 20 (yellow) bosons obtained using the ID approach.  The dotted line is the result calculated using HEOM with the ESPRIT decomposition.}
    \label{fig:pop-subohmic}
\end{figure}

In FIG. \ref{fig:bcf-subohmic} and \ref{fig:pop-subohmic}, we present the BCF and population dynamics for the sub-Ohmic case ($s=0.25$), respectively.  The temperature is set to $50\;\text{K}$.  We point out that the sub-Ohmic SD is critical at finite temperature because the integrand of the BCF exhibits a singularity at $\omega = 0$ caused by the Bose-Einstein distribution, which disappears in the zero temperature limit.  The cutoff time and cutoff frequency for the ID approach are $T = 1000\;\text{fs}$ and $\Omega = 200\;\mathrm{cm}^{-1}$, respectively.  The BCF is approximated using $20$ sample points, which allow the population dynamics to converge with the ID approach (see FIG. \ref{fig:pop-subohmic}).  The ID approach can accurately approximate the BCF over the entire time range, whereas the other methods fail to achieve comparable accuracy.  To investigate the behaviour with the number of sample points increasing, we also show the BCF approximated with 500 sample points in FIG. \ref{fig:bcf-subohmic2}.  Even with 500 points, MDM and BSDO approaches cannot give accurate results because of the singularity at $\omega=0$.  This implies that the methods based on orthogonal polynomials, such as T-TEDOPA,\cite{TamascelliEtAl2019PRL,RivaEtAl2023PRBa} fail for this class of QNSDs.  Conversely, the ID approach does not suffer from the singularity because it can exploit the temporal information enclosed 
in the BCF through the NNLS procedure.  
The LD approach provides a good approximation of the SD only when using a very large number of points -- over 500 are needed to match the accuracy that the ID method achieves with just 20 points.
In FIG. \ref{fig:pop-subohmic}, we can observe that the dynamics rapidly converges as the number of sample points increase, and the result with 13 points is already very accurate.  The BCF calculated using the ID approach with $20$ vibrations agrees with the HEOM result.  

\section{Discretization of structured spectral densities: Electron transfer in biological systems\label{sec:result2}}
Complex molecular systems are often 
characterized by structured spectral densities whose rich features can strongly impact their dynamical behaviour.  In order to test the effectiveness of the current method beyond the simple unstructured Ohmic spectral densities, we apply it to determine
a low-rank discrete model 
of the electron-transfer process in plant cryptochromes from \textit{Arabidopsis thaliana} (AtCry).
The SD of this process has been obtained from QM/MM simulations with the constrained DFT approach. \cite{RezacEtAl2012JCTC,FirminoEtAl2016PCCP,Mendive-TapiaEtAl2018JPCB}.  Using the discretized bath parameters, we also calculate the electron transfer dynamics, and compare it with previous results obtained with the ML-MCTDH method.\cite{Mendive-TapiaEtAl2018JPCB}

Electron transfer in protein can be modelled using a two-state donor-acceptor model.  The Hamiltonian is readily written as
\begin{align}
    H &= H_\mathrm{D}\ketbra{D}+H_\mathrm{A} \ketbra{A} +H_\mathrm{DA}(\ketbra{D}{A}+\ketbra{A}{D}) \notag\\
    &+ \sum_{k} \frac{\omega_k}{2}\qty(p_k^2+x_k^2) + \sum_{k} g_k x_k\qty(\ketbra{D}-\ketbra{A}),
\end{align}
where $\ket{D}$ and $\ket{A}$ denote the donor and acceptor states, respectively.  $H_\mathrm{D}$ and $H_\mathrm{A}$ represent the energies of each state, while $H_\mathrm{DA}$ denotes the electronic coupling between the donor and acceptor states. Here, $x_k$, $p_k$, and $\omega_k$ are the position, momentum, and frequency of the $k$th vibrational mode, respectively. The term $g_k$ represents the vibrational coupling strength of the $k$th mode, which is related to the displacement $g_k/\omega_k$. This model Hamiltonian assumes that the harmonic potentials in both electronic states have the same frequency, position, and momentum coordinates.
We use the parameters $H_\mathrm{D}=0.0$, $H_\mathrm{A} = -511.63$ and $H_\mathrm{DA}=19.666$ meV, adopted from Ref. \cite{Mendive-TapiaEtAl2018JPCB}. 
In our model, the reorganization energy is set to $1.34\;\mathrm{eV}$.

Before discretization is performed, we smooth the noise of the SD by employing a smoothing spline to improve the numerical stability of ID.  We then apply an interpolation algorithm to the smoothed data to obtain an analytical expression for the spectral density.  In this study, we use the AAA algorithm.  This expression can be used to form the matrix $\mathbf{f}$ and calculate the BCF by numerically integrating Eq. \eqref{eq:fdt1}.  
Due to the low reliability of the low frequency part of the spectral density, we set the spectral density to zero for $\omega<1\;\mathrm{cm}^{-1}$, which is a common practice in the field of electron transfer dynamics\cite{Mendive-TapiaEtAl2018JPCB}.  The spectral density is shown in FIG. \ref{fig:sbeta250} (a).

In this section, comparison is only made against the BSDO approach, since it outperforms LD and MDM approaches even in the case of simple SDs, as we have seen in the previous section.
For both the ID and BSDO approaches, we consider the frequency interval $[0,4000]\;\mathrm{cm}^{-1}$ unless otherwise specified.  The initial grids for the ID approach comprise 1000 points in time domain and 2000 points in frequency domain for all the results in this section.

\subsection{Zero temperature}
\begin{figure}[t]
    \centering
    \includegraphics[width=8.4cm]{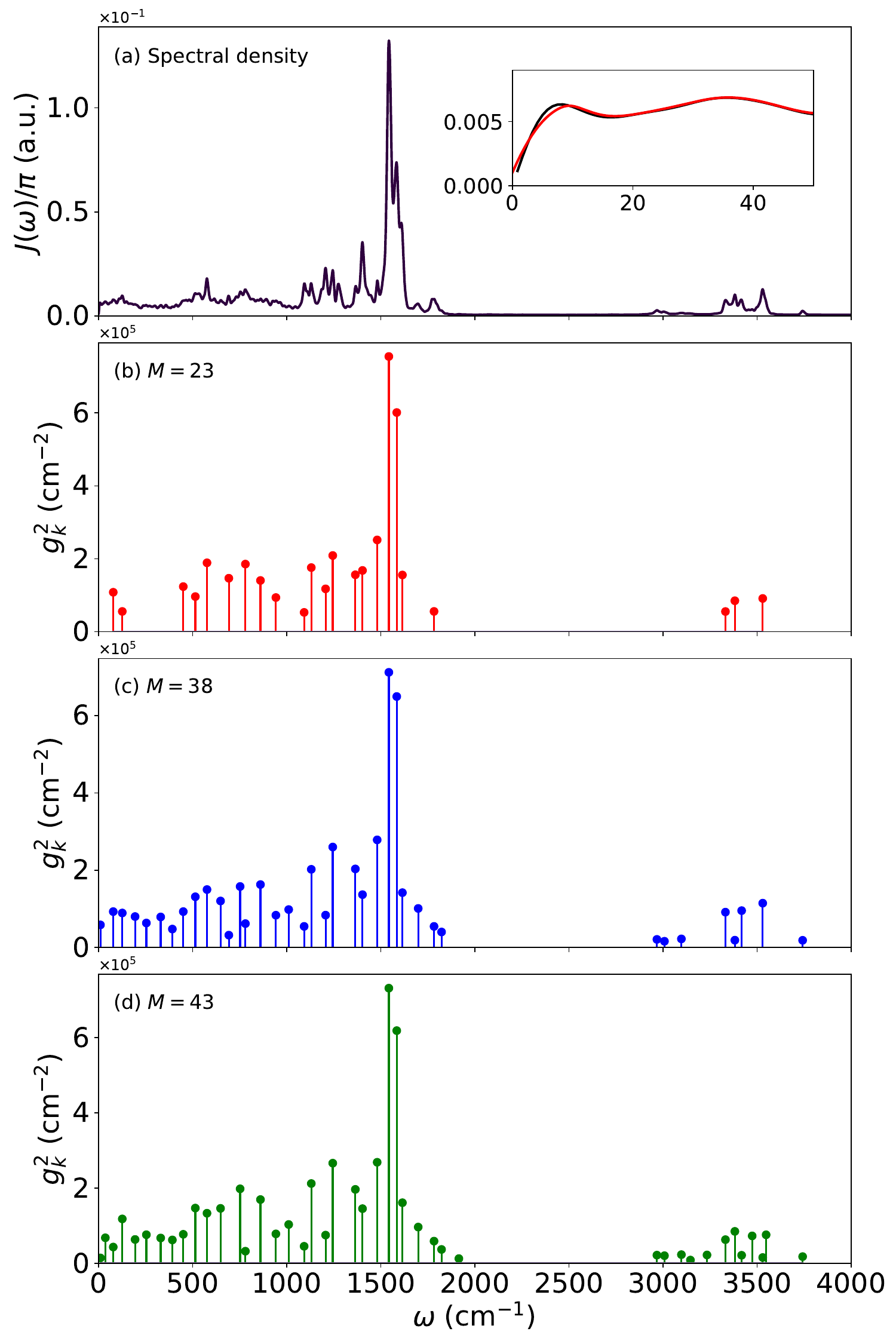}
    \caption{(a) Spectral density $J(\omega)/\pi$ for the cryptochromes (black), coefficients $g_k^2$ and their corresponding frequencies $\omega_k$ of (b) 23, (c) 38 and (d) 43 vibrations obtained using the ID approach with $T=250\;\mathrm{fs}$.  The inset in (a) shows the low-frequency region of the spectral density, including the original data (black) and the fitted data (red) obtained using the AAA algorithm.}
    \label{fig:sbeta250}
\end{figure}

\begin{figure}[t]
    \centering
    \includegraphics[width=8.4cm]{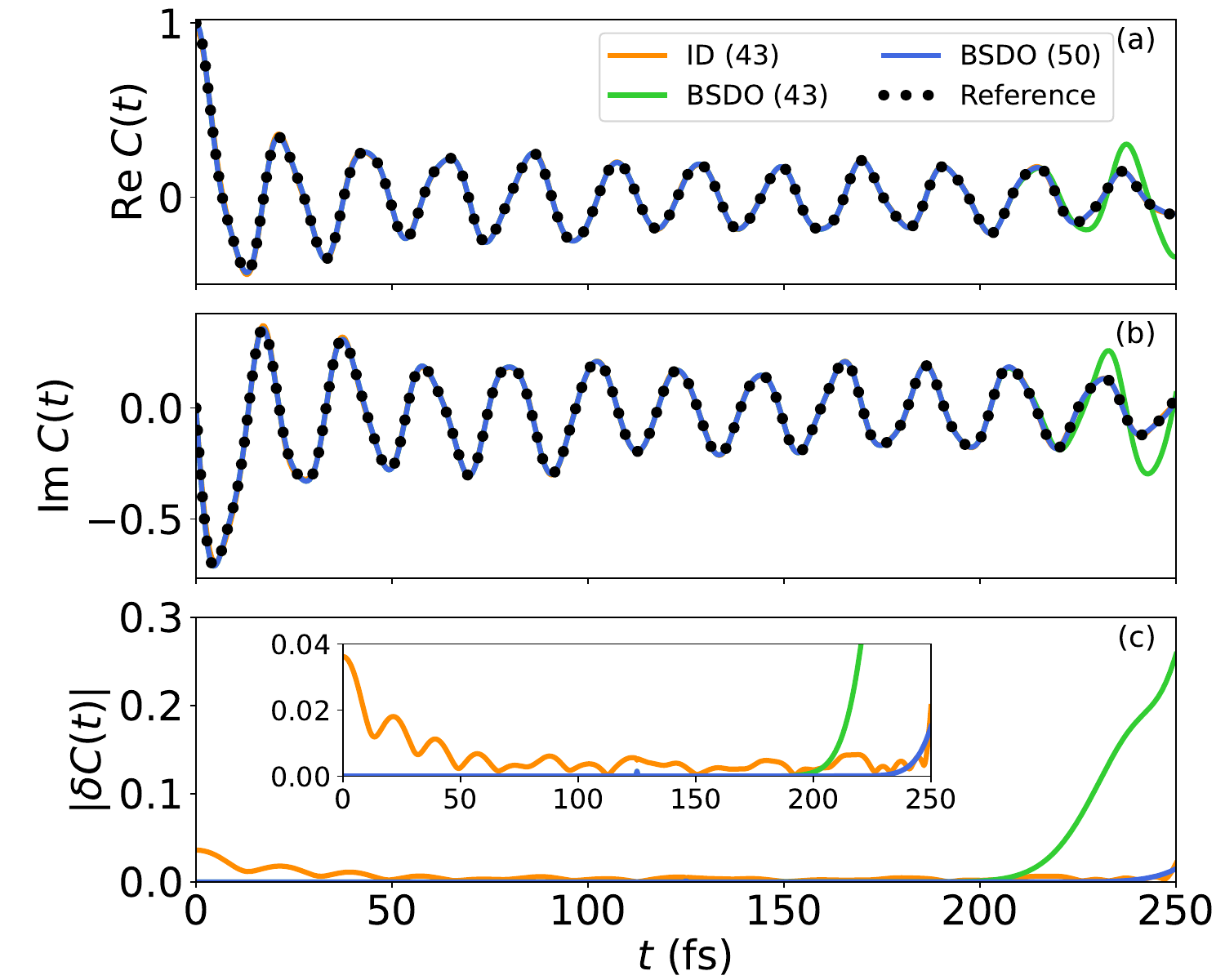}
    \caption{(a) Real part, (b) imaginary part and (c) absolute error of the bath correlation function for the cryptochromes approximated using the ID (red), BSDO with 43 sample points (green) and BSDO with 50 sample points.  The absolute error is defined as $|\delta C(t)|=|C_\mathrm{approx}(t)-C(t)|$.  Black dotted lines are the references.  All the results are normalized to the absolute value of $C(0)$.}
    \label{fig:bcf250}
\end{figure}

\begin{figure}[t]
    \centering
    \includegraphics[width=8.4cm]{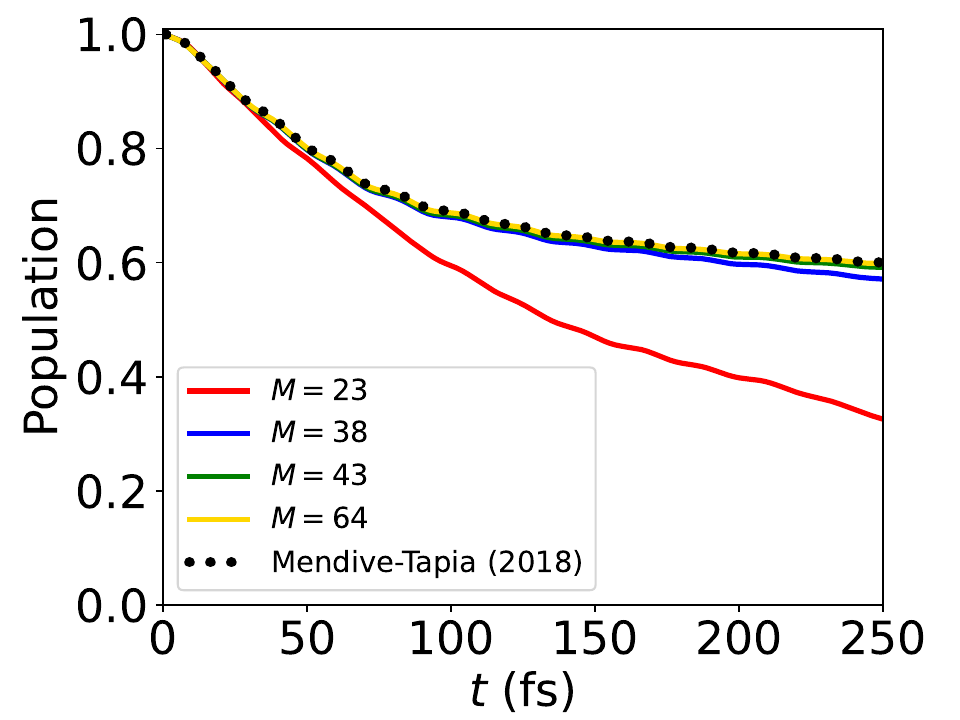}
    \caption{Population dynamics of the donor state at $0$ K calculated using 23 (red), 38 (blue), 43 (green) and 64 (yellow) vibrations obtained using the ID approach.  The dotted line is the dynamics calculated using ML-MCTDH, taken from Mendive-Tapia et. al. (2018)\cite{Mendive-TapiaEtAl2018JPCB}.}
    \label{fig:pop250}
\end{figure}

First, we consider the zero-temperature case.  In FIG. \ref{fig:sbeta250}, we present the sample frequencies $\omega_k$ and corresponding coefficients $g_k^2$ obtained using the ID approach with a cutoff time of $T=250\;\mathrm{fs}$. 
We observe that the ID approach can effectively capture the characteristic features of the SD.
The population dynamics are shown in FIG. \ref{fig:pop250}.  The dynamics converge rapidly as the number of sample points increases.
The result obtained with 43 vibrations is already in good agreement with the ML-MCTDH result obtained with an equispaced discretization with at least 256 frequencies.\cite{Mendive-TapiaEtAl2018JPCB}  Thus, the ID approach can provide an accurate result using five times fewer sample points.  FIG. \ref{fig:bcf250} displays the corresponding BCF with 43 sample points.  When we apply the BSDO with the same number of sample points, it yields a more accurate result at early times but starts to deviate from the reference around $t=200\;\mathrm{fs}$, and requires 50 sample points to obtain an accurate BCF up to $250\;\mathrm{fs}$.

\begin{figure}[t]
    \centering
    \includegraphics[width=8.4cm]{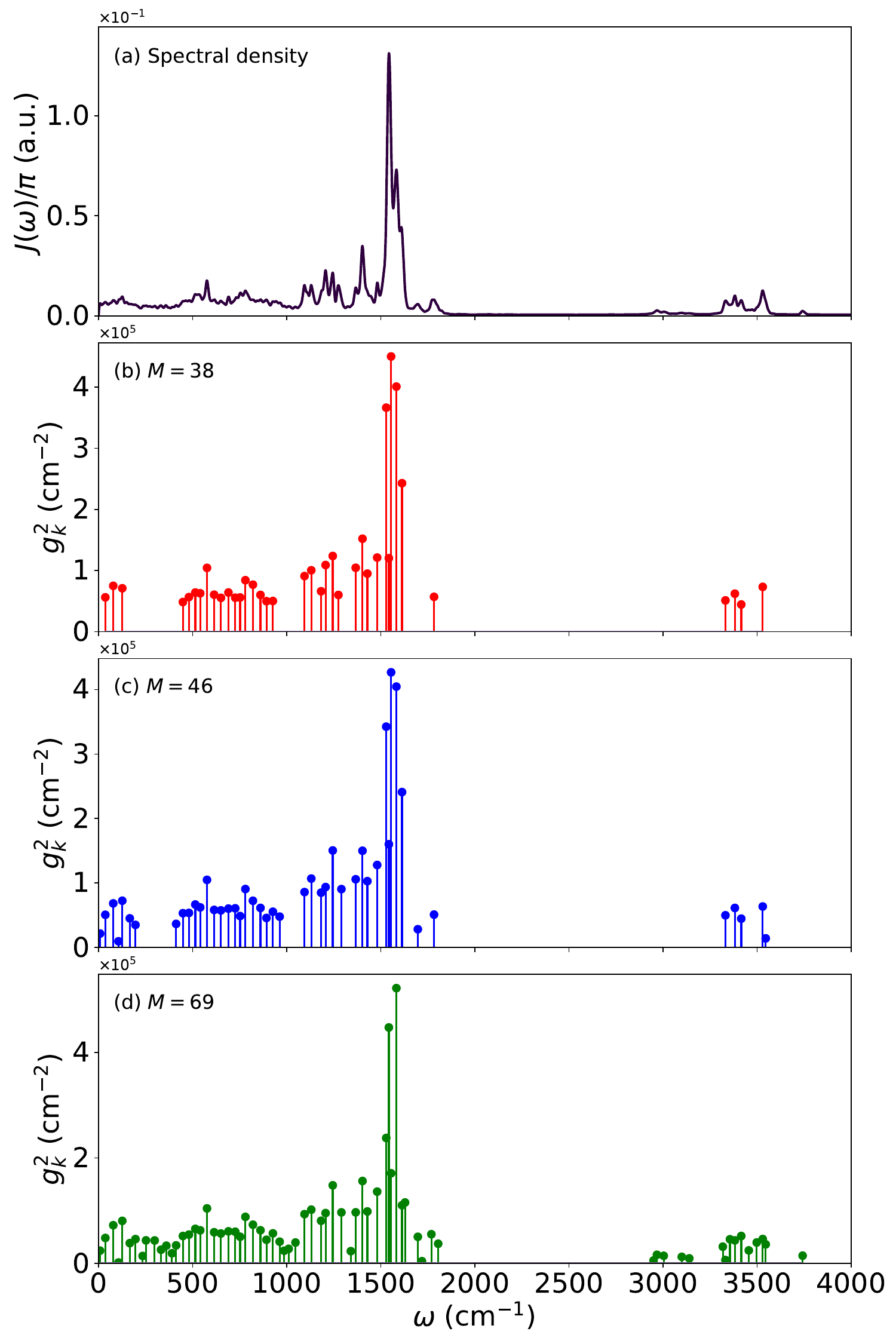}
    \caption{(a) Spectral density $J(\omega)/\pi$ for the cryptochromes (black), coefficients $g_k^2$ and their corresponding frequencies $\omega_k$ of (b) 38, (c) 46 and (d) 69 vibrations using the ID apporach with $T=500\;\mathrm{fs}$.}
    \label{fig:sbeta500}
\end{figure}

\begin{figure}[t]
    \centering
    \includegraphics[width=8.4cm]{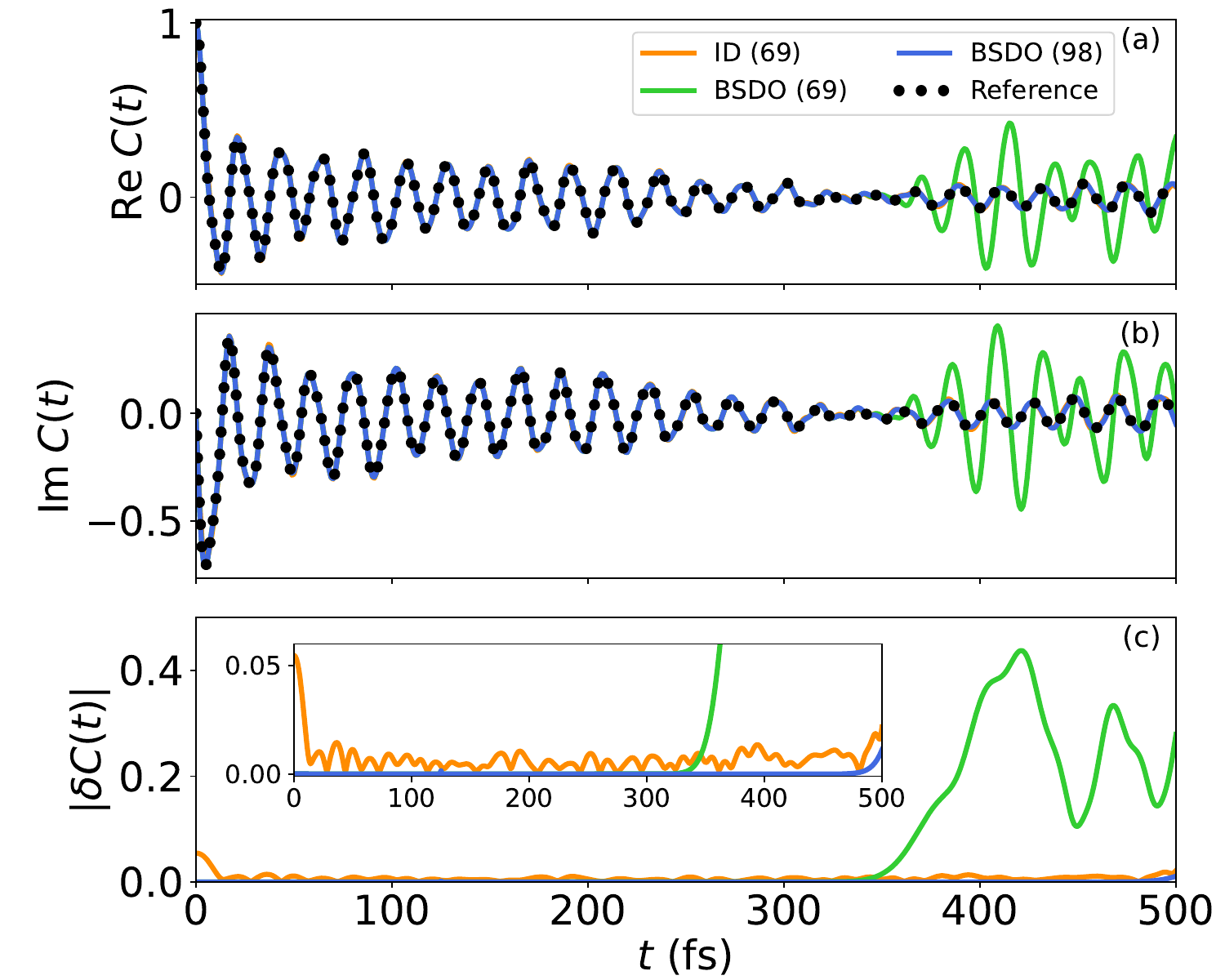}
    \caption{(a) Real part, (b) imaginary part and (c) error of the BCF at 0 K for the cryptochromes approximated using the ID (red), BSDO with 69 sample points (green) and BSDO with 98 sample points (blue).  The error is defined as $\delta C(t)=C_\mathrm{approx}(t)-C(t)$.  Black dotted lines are the references.  All the results are normalized to the absolute value of $C(0)$.}
    \label{fig:bcf500}
\end{figure}

\begin{figure}[h]
    \centering
    \includegraphics[width=8.4cm]{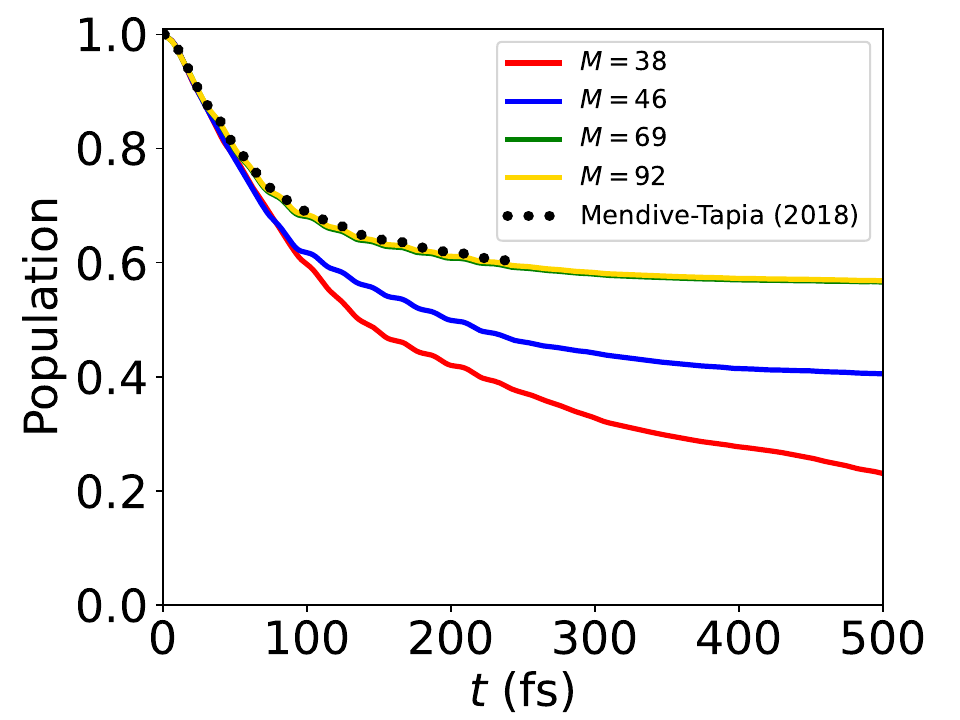}
    \caption{Population dynamics of the donor state at 0 K calculated using 38 (red), 46 (blue), 69 (green) and 92 (yellow) vibrations obtained using the ID approach.  The dotted line is the dynamics calculated using ML-MCTDH, taken from Mendive-Tapia et. al. (2018).\cite{Mendive-TapiaEtAl2018JPCB}}
    \label{fig:pop500}
\end{figure}

Next, we consider longer time dynamics to demonstrate the flexibility and robustness of the ID approach.  FIG. \ref{fig:sbeta500} displays the frequencies $\omega_k$ and the corresponding coefficients $g_k^2$ obtained using the ID approach with $T=500\;\mathrm{fs}$.  Compared to the discretization obtained using $T=250\;\mathrm{fs}$, in this case the sample frequencies are more densely distributed in the low-frequency domain which is expected in order to capture long-time behaviour.  As shown in FIG. \ref{fig:pop500}, using the ID approach, the population dynamics converges with 69 vibrational modes.  The corresponding BCF is shown in FIG. \ref{fig:bcf500}.  The BSDO does not yield an accurate BCF over the entire time domain with 69 points and requires nearly 100 sample points to achieve convergence.
According to the Lieb-Robinson bound, the number of necessary bath modes required by the BSDO approach can be predicted to increase proportionally with time $t$, that is, $M \propto t$,\cite{WoodsPlenio2016JMP} which agrees well with our current results. In contrast, the ID approach is more robust
can capture the long-time behaviour with a smaller number of sample points. 

However, we observe that achieving very high accuracy using the ID method is more challenging compared to the BSDO approach. Indeed, our results suggest that the overall accuracy of the ID does not increase significantly despite a greater number of sample points (refer to the SM). Nevertheless, ID consistently demonstrates to be practically reliable for dynamical simulations.

\subsection{Finite temperature}
\begin{figure}[t]
    \centering
    \includegraphics[width=8.4cm]{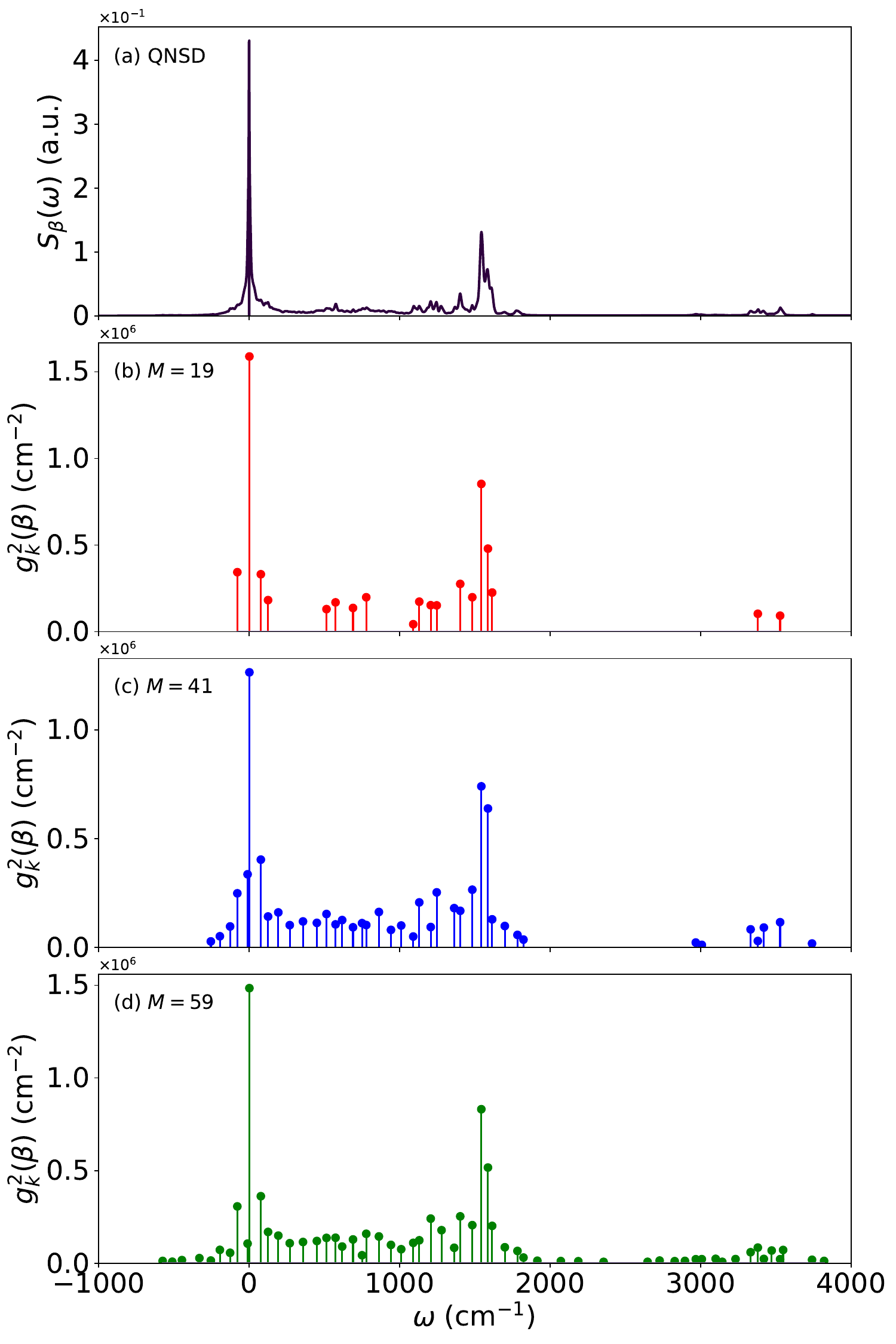}
    \caption{(a) Quantum noise spectral density $S_\beta(\omega)$ for the cryptochromes at 300 K (black), coefficients $g_k^2(\beta)$ and their corresponding frequencies $\omega_k$ of (b) 11, (c) 41 and (d) 59 vibrations obtained using the ID approach.}
    \label{fig:sbeta300K}
\end{figure}

\begin{figure}[h]
    \centering
    \includegraphics[width=8.4cm]{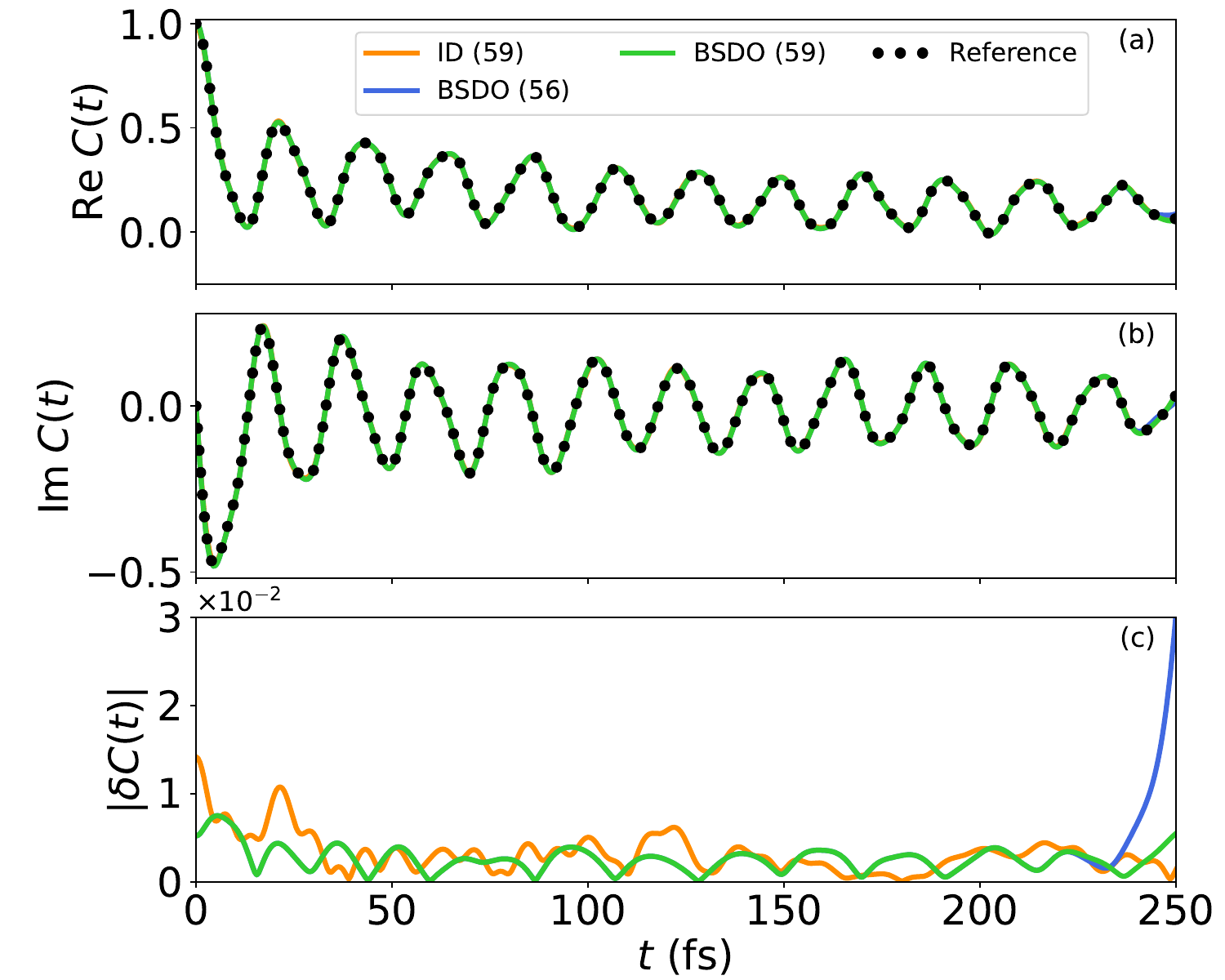}
    \caption{(a) Real part, (b) imaginary part and (c) absolute error of the BCF at 300 K for the cryptochromes approximated using the ID (red), BSDO with 59 sample points (green) and BSDO with 56 sample points (blue).  The absolute error is defined as $|\delta C(t)|=|C_\mathrm{approx}(t)-C(t)|$.  Black dotted lines are the references.  All the results are normalized to the absolute value of $C(0)$.}
    \label{fig:bcf300K}
\end{figure}

\begin{figure}[h]
    \centering
    \includegraphics[width=8.4cm]{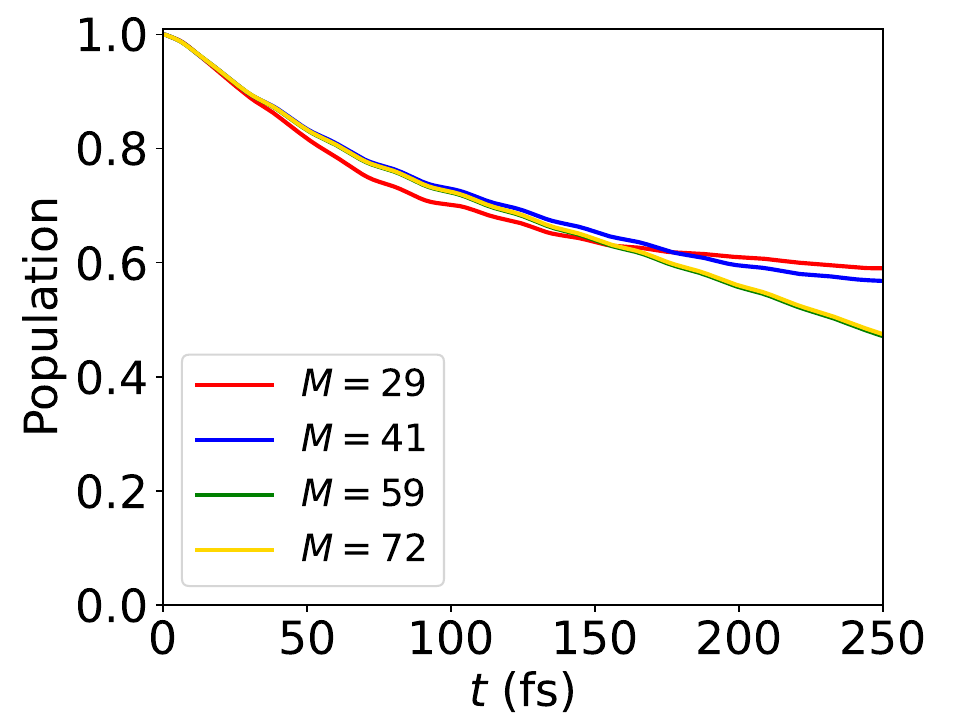}
    \caption{Population dynamics of the donor state at 300 K calculated using 29 (red), 41 (blue), 59 (green) and 72 (yellow) vibrations.}
    \label{fig:pop300K}
\end{figure}

Lastly, we test the ID and BSDO approaches for a finite-temperature case at $300\;\text{K}$.  The QNSD is shown in FIG. \ref{fig:sbeta300K} (a).  The frequencies $\omega_k$ and coefficients $g_k^2$ obtained using the ID approach with different sample points are also presented in FIG. \ref{fig:sbeta300K} (b)-(d).  The QNSD has a sharp peak around $\omega = 0$ and $\omega = 1500$, and the ID effectively detects these significant frequencies.
FIG. \ref{fig:bcf300K} and \ref{fig:pop300K} display the BCF and the population dynamics, respectively. Using the ID approach, the population dynamics converges with 59 vibrational modes. 

BCFs approximated using the ID and BSDO approaches with 59 sample points exhibit comparable accuracy. However, we found that BSDO results are highly sensitive to the spectral density behavior in the low-frequency region. As previously demonstrated, BSDO can encounter convergence issues with sub-Ohmic spectral densities due to the zero-frequency singularity. At finite temperatures, this issue 
can originate from the \(\coth(\beta\omega/2)\) factor, leading to divergence. 
In our application, setting the components of \(J(\omega)\) below 1 cm\(^{-1}\) to zero enabled effective use of BSDO. Additionally, the frequency interval for BSDO was carefully selected as \([-700, 4000]\; \mathrm{cm}^{-1}\), as deviations from this range resulted in lower accuracy. By contrast, the ID methodology remains entirely unaffected by such factors. 

\begin{figure}[h]
    \centering
    \includegraphics[width=8.4cm]{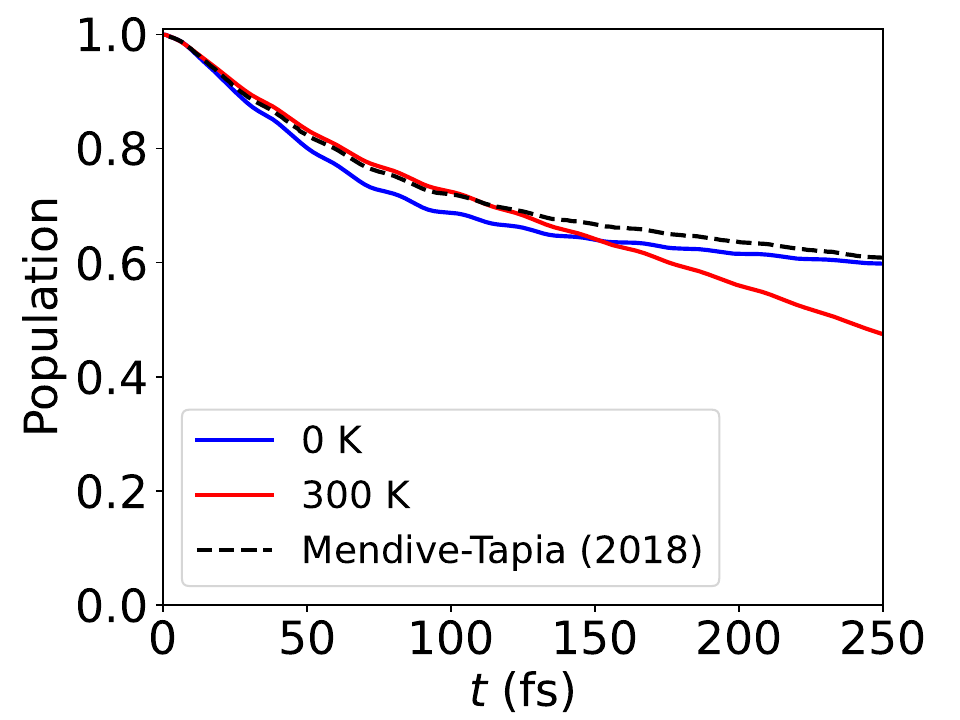}
    \caption{Population dynamics of the donor state at 0K (blue) and 300 K (red).  The converged results are presented.  The dashed line is the dynamics calculated using ML-MCTDH combined with the random phase thermal wave function approach\cite{NestKosloff2007JCP}, taken from Mendive-Tapia et. al. (2018).\cite{Mendive-TapiaEtAl2018JPCB}
    }
    \label{fig:pop_0K300K}
\end{figure}

In FIG \ref{fig:pop_0K300K}, we compare our population dynamics with those calculated in a previous study based on ML-MCTDH and the random phase thermal wave function approach. The populations behave similarly up to $t \approx 100, \mathrm{fs}$, after which they start to deviate from each other. Considering that we do not impose any approximations on the dynamics other than the tensor-train ansatz, we conclude that in this system the random phase thermal wave function approach does not properly account for thermal effects.

\section{Discussion and Conclusion \label{sec:conclusion}}

We have proposed a new approach to construct an effective discrete bath Hamiltonian by approximating the BCF using ID theory.  Through a comparative analysis with other methodologies, we have shown that the ID approach can both reliably and accurately approximate the BCF using fewer sample points than the alternative techniques.

Two key characteristics distinguish the ID approach from earlier methods. Firstly, it leverages the BCF-QNSD relation and its low-rank decomposition. Additionally, utilizing ID enables the automatic selection of significant frequencies and provides a discrete bath Hamiltonian that incorporates all requisite information for conducting dynamical simulations at a specified level of accuracy.
The second characteristic is the emphasis on approximating the BCF instead of directly discretizing the spectral density.
The BCF is the key element determining the system's dynamics. Accurately approximating it is essential for capturing the system’s behavior effectively.
Moreover, our method can validate the convergence with respect to the number of sample vibrations by directly assessing the approximation error of the BCF. 
Modeling ET processes at finite temperature in biological systems is
a rather complex problems and can require several hundred  vibrations.\cite{Borrelli2018CP} Here, we have demonstrated that our approach can provide converged results with a few tens of sample frequencies.

Through various examples, we have shown that the present method outperforms previous methods in the majority of cases. 
In some cases, the BSDO method can achieve accuracy comparable to the ID approach.
However, for the BSDO method to be effectively utilized, it is essential to properly select the frequency range, which necessitates multiple adjustments through iterative trial-and-error procedures.  Furthermore, the BSDO methodology encounters an issue with sub-Ohmic QNSDs due to the Gaussian quadrature weight becoming unbounded at $\omega = 0$. On the other hand, the ID method is more robust and free from these difficulties. 
Conversely, when the QNSD are not sub-Ohmic, the ID approximation for long-time dynamics can be less accurate than the BSDO. 

Finally, we have shown that by integrating this approach with the tensor-train formalism, we can efficiently investigate the quantum dynamics of systems interacting with complex non-Markovian environments. This combination provides reliable and robust computational framework for exploring the dynamics of realistic chemico-physical systems. Although our current study focuses only on bosonic baths, the same techniques can also be effectively applied to fermionic baths.  Further research is currently being conducted in this area.

\begin{suppinfo}
The codes used in our simulations are available at the address: \texttt{https://github.com/htkhsh/QFiND}.
The supporting information provides additional data.
\end{suppinfo}

\begin{acknowledgement}

This work has been partially supported by the Spoke 7 "Materials and Molecular Sciences" of ICSC – Centro Nazionale di Ricerca in High-Performance Computing, Big Data and Quantum Computing, funded by the European Union – NextGenerationEU.
The authors further acknowledge
the University of Torino for the local research funding Grant No.
BORR-RILO-22-01.

R.B. acknowledges the research project “nuovi Concetti, mAteriali e tecnologie per l’iNtegrazione del fotoVoltAico negli edifici in uno scenario di generazione diffuSa” [CANVAS], funded by the Italian Ministry of the Environment and the Energy Security, through the Research Fund for the Italian Electrical System (type-A call, published on G.U.R.I. n. 192 on 18-08-2022)
 
\end{acknowledgement}

\appendix
\section{Alternative discretization approaches}
We briefly review the discretization methods used for comparison. The problem we consider here is the discretization of $S_\beta(\omega)$ in the frequency interval $[-\Omega, \Omega] \backslash \{0\}$ using $M$ sample frequencies, $\omega_k\;(k=1, \dots, M)$ and the corresponding coefficients $g_k(\beta)$.  Some of the methods have been originally developed for the zero-temperature case, and we have adapted them for the finite-temperature case.  Note that the order of the frequencies of $\omega_k$ is not considered here.

\subsection{Logarithmic discretization}
The logarithmic discretization is suitable for characterizing the low-frequency domain of the spectral density.  We apply this method to positive and negative frequency domains separetely.  Thus we divide the frequency intervals $[-\Omega,0)$ and $(0,\Omega]$ into $M-2$ domains $[-\Lambda^{-(k-1)}\Omega,-\Lambda^{-k}\Omega]\;(k=1, \dots, M/2-1)$ and $[\Lambda^{-(k-M/2)}\Omega, \Lambda^{-(k-1-M/2)}\Omega]\;(k=M/2+1, \dots, M)$, respectively.  Here we assume that $M$ is even.  The parameters $\omega_k$ and $g_k(\beta)$ can be obtained as
\begin{equation}
    g_k^2(\beta) = \left\{
    \begin{alignedat}{2} 
        &\int_{-\Lambda^{-(k-1)}\Omega}^{-\Lambda^{-k}\Omega}  \dd \omega \; S_\beta(\omega) & \quad (k=1, \dots, \frac{M}{2})\\
        &\int_{\Lambda^{-k-M/2}\Omega}^{\Lambda^{-(k-1-M/2)}\Omega}  \dd \omega \; S_\beta(\omega) & \quad (k=\frac{M}{2}+1  , \dots, M)
    \end{alignedat}
    \right.
\end{equation}
and
\begin{equation}
    \omega_k = \left\{
    \begin{alignedat}{2} 
        &g_k^{-2}(\beta)\int_{-\Lambda^{-(k-1)}\Omega}^{-\Lambda^{-k}\Omega}  \dd \omega \; S_\beta(\omega)\omega & \quad (k=1, \dots, \frac{M}{2})\\
        &g_k^{-2}(\beta)\int_{\Lambda^{-k-M/2}\Omega}^{\Lambda^{-(k-1-M/2)}\Omega}  \dd \omega \; S_\beta(\omega)\omega & \quad (k=\frac{M}{2}+1  , \dots, M)
    \end{alignedat}
    \right.
\end{equation}
In this study, we fix the discretization parameter to $\Lambda=1.1$, which is used in Ref. \cite{WangEtAl2016JCP}.

\subsection{Mode density method}
The method by Walters et al.\cite{WaltersEtAl2017JCC}, here referred to as the mode density method (MDM), divides the frequency axis into domains with equal weight based on the mode density $\rho(\omega)$ so that each domain has equal renormalization energy.  For finite temperatures, we first apply the MDM to the spectral density and obtain the negative frequencies by doubling the obtained sample frequencies.  Here, we assume that $M$ is even.

Therefore, the $k$th sample point should satisfy 
\begin{equation}
    \int_0^{\omega_k} \dd \omega\; \rho(\omega)=k-\frac{1}{2}, \;\;(k=1,\cdots,\frac{M}{2})
\end{equation}
and 
\begin{equation}
    \int_0^{\infty} \dd \omega\; \rho(\omega)=\frac{M}{2}.
\end{equation}

Here we employ the mode density
\begin{equation}
    \rho(\omega) = \Gamma^{-1} \frac{J(\omega)}{\omega}
\end{equation}
where $\Gamma$ is the normalization factor
\begin{equation}
    \Gamma = \qty(\frac{M}{2})^{-1} \int_0^{\infty}\dd \omega\; \frac{J(\omega)}{\omega}.
    \label{eq:}
\end{equation}
Then sample frequencies $\omega_k$ in the positive domain are determined by solving the nonlinear problem
\begin{equation}
    \Gamma^{-1} \int_0^{\omega_k}\dd \omega\; \frac{J(\omega)}{\omega} =k-\frac{1}{2}, \;\;(k=1,\cdots,\frac{M}{2}),
\end{equation}
and the negative frequencies are $\omega_{k+M/2}=-\omega_k$.
The coefficients are obtained as
\begin{equation}
    g_k^2 = \frac{S_\beta(\omega_k)}{\rho(\omega_k)}, \;\;(k=1,\cdots,M).
\end{equation}

\subsection{Gauss quadrature with the BSDO weight}
BSDO method uses Gauss quadrature to numerically integrate Eq. \eqref{eq:fdt1}\cite{deVegaEtAl2015PRB,WoodsPlenio2016JMP}.  The unique feature of the BSDO method is that it uses the SD as the weight of a quadrature, enabling to construct an efficient set of polynomial interpolants.  In this method, we consider a more general frequency interval $[\Omega_\mathrm{min},\Omega_\mathrm{max}]$.  Note that the choice of $\Omega_\mathrm{min}$ and $\Omega_\mathrm{max}$ affects the accuracy and convergence of the approximation, and it is desirable that the interval covers the entire frequency range of the QNSD i.e. $S_\beta(\omega)$ should be small enough for $\omega<\Omega_\mathrm{min}$ and $\omega>\Omega_\mathrm{max}$.

We first introduce the so-called hybridization function
\begin{equation}
    \label{eq:lambda}
    \Lambda(z)=\int_{\Omega_\mathrm{min}}^{\Omega_\mathrm{max}} \dd \omega \;\frac{S_\beta(\omega)}{z-\omega}, \quad z \in \mathbb{C}.
\end{equation}
By the Sokhotski-Plemelj theorem, Eq. \eqref{eq:lambda} implies
\begin{equation}
    S_\beta(\omega)=-\frac{1}{\pi} \lim_{\epsilon\rightarrow 0^+}\operatorname{Im} \Lambda(\omega+i \epsilon).
\end{equation}
If $\Lambda(z)$ is approximated with a sum of rational functions as
\begin{equation}
    \label{eq:lambda_disc}
    \Lambda(z)\approx\sum_{k=1}^M\frac{g_k^2(\beta)}{z-\omega_k},
\end{equation}
the BCF-QNSD relation can be expressed 
\begin{align}
    C(t) &= -\frac{1}{\pi} \operatorname{Im}\int_{-\infty}^\infty \dd\omega\; \lim_{\epsilon\rightarrow 0^+} \Lambda(\omega+i\epsilon) \mathrm{e}^{-i\omega t}\\
    &=\sum_{k=1}^M g_k^2(\beta)\mathrm{e}^{-i\omega_k t}.
\end{align}
Thus we consider expressing $\Lambda(z)$ in the form of Eq. \eqref{eq:lambda_disc} using a Gauss quadrature.

To achieve this, we express the integral Eq. \eqref{eq:lambda} in terms of the product of a weight function $w(\omega)\geq 0$ and a function $h(\omega,z)$,
\begin{equation}
    \Lambda(z)=\int_{\Omega_\mathrm{mix}}^{\Omega_\mathrm{max}} \dd \omega \;\frac{S_\beta(\omega)}{z-\omega}=\int_{\Omega_\mathrm{mix}}^{\Omega_\mathrm{max}} \dd \omega\; w(\omega) h(\omega, z).
\end{equation}
Now we consider a polynomial interpolant $h_M(\omega,z)$ of $h(\omega,z)$ with degree $M-1$
\begin{align}
    h(\omega,z) & = h_M(\omega,z)+r_M(\omega, z) \\
    h_M(\omega,z) & =\sum_{k=1}^M h(\omega_k,z) l_k(\omega), \quad l_k(\omega_l)=\delta_{kl}
\end{align}
where $l_k(\omega)$ can be defined as the $(M-1)$-th order polynomial $l_k(\omega)=\prod_{k \neq l}(\omega-\omega_k) / \prod_{l \neq k}(\omega_k-\omega_l)$ and $r_M(\omega,z)$ is a remainder. 
If choosing the $M$ node points $\omega_k$,
\begin{equation}
    \Lambda(z) =\int_{\Omega_\mathrm{mix}}^{\Omega_\mathrm{max}} \dd \omega\; w(\omega) h(\omega, z)\approx\sum_{k=1}^M W_k h(\omega_k, z)
\end{equation}
with $W_k =\int_{\Omega_\mathrm{mix}}^{\Omega_\mathrm{max}} \dd\omega \; w(\omega) l_k(\omega)$, which are referred to as Christoffel weights.

In the BSDO approach, the QNSD (originally the SD) is chosen as the weight of the polynomials i.e. $w(\omega)=S_\beta(\omega)$ and $h(\omega,z)=\frac{1}{z-\omega}$.  The nodes can be determined as the roots of the orthogonal polynomial $p_M(\omega)$ of degree $M$, obeying 
\begin{equation}
    \int_{\Omega_\mathrm{mix}}^{\Omega_\mathrm{max}} \dd \omega \;S_\beta(\omega) p_k(\omega) p_l(\omega)=\delta_{k l}.
\end{equation}
Such polynomials can be generated using the recurrence relation
\begin{equation}
    p_{k+1}(\omega) =(\omega-\alpha_k) p_k(\omega)-\eta_k p_{k-1}(\omega),\quad k=0, \dots, M-1
\end{equation}
where $p_0(\omega) =1, p_{-1}(\omega)=0$.  $\alpha_k$ and $\eta_k$ are the recurrence coefficients, which can be obtained, by applying the Lanczos algorithm,\cite{GraggHarrod1984NM,Gautschi1994ATMS,Gautschi2005JoCaAM}  as the elements of a tridiagonal matrix
\begin{equation}
    \mathbf{A}_M = \qty(\begin{array}{cccc}
    \alpha_0 & \sqrt{\eta_1} & 0 & \ldots \\
    \sqrt{\eta_1} & \alpha_1 & \sqrt{\eta_2} & \ddots \\
    0 & \sqrt{\eta_2} & \alpha_2 & \ddots \\
    \vdots & \ddots & \ddots & \ddots
    \end{array}) \in \mathbb{R}^{M \times M}.
\end{equation}
By performing an eigenvalue decomposition, 
$\mathbf{A}_M = \mathbf{U}\mathrm{diag}(\omega_1,\dots,\omega_M)\mathbf{U}^{\mathrm{T}}$, we can obtain the nodes of $p_M$, leading to sample frequencies $\omega_k$, as the eigenvalues .  The Christoffel weights are given by the square of the first element of each eigenvector
\begin{equation}
    g_k^2(\beta)=\Delta_w [\mathbf{U}(1,k)]^2 
\end{equation}
where $\Delta_w=\int_{\Omega_\mathrm{mix}}^{\Omega_\mathrm{max}} \dd \omega\; S_\beta(\omega)$ is the normalization factor.  

It is worth remarking that from the matrix $\mathbf{A}_M$, we can construct the chain representation of the system,\cite{PriorEtAl2010PRL, deVegaEtAl2015PRB} which is expressed with the Hamiltonian
\begin{align}
    H_\mathrm{chain} &= H_\mathrm{S} + \sqrt{\Delta_w} V_\mathrm{SB} (b_0^\dag + b_0) + \sum_{k=0}^{M-1} \alpha_{k} b_k^\dagger b_k \notag\\
    &\hspace{10pt} + \sum_{k=0}^{M-2} \sqrt{\eta_k} (b_k^\dag b_{k+1} + b_k b_{k+1}^\dag).
\end{align} 
where $\qty{b_k,b_k^\dag}$ is a set of bosonic creation and annihilation operators other than the original.
This formulation is nothing else but the T-TEDOPA formulation.\cite{TamascelliEtAl2019PRL}

\section{Tensor-train format}
To solve the time-dependent Schrödinger equation Eq. \eqref{eq:tseq} with the Hamiltonian Eq. \eqref{eq:tfdgenericham}, it is inevitable to utilize effective numerical strategies capable of accurately treating a large number of dynamical variables.  In our approach, the TT decomposition, the simplest form of tensor networks, has been adopted.  We outline the fundamental concepts of TT decomposition here and show its application in solving the thermal Schrödinger equation.   The reader is referred to the original articles for a detailed analysis of the TT theory\cite{Oseledets2011SJSC}.

Let us consider a generic state $\ket{\psi}$ of a $N$ dimensional quantum system having the form
\begin{equation}
    \ket{\psi}=\sum_{i_1, i_2, \ldots, i_N} A(i_1, \ldots, i_N)\ket{i_1}\ket{i_2}\cdots\ket{i_N} .
\end{equation}
where $\ket{i_k}$ labels the basis states of the $k$-th dynamical variable with $n_k$ states, and the elements $A(i_1, \ldots, i_N)$ are complex numbers labeled by $N$ indices and represent a tensor of order $N$.  With the help of the TT format, each element $A(i_1, \ldots, i_N)$ of the tensor $A$ is approximated as
\begin{equation}
	\label{eq:ttformat}
	A(i_1,...,i_N) \approx \sum_{\alpha_0, \cdots,\alpha_N=1}^{r_0,\ldots,r_N} G^{(1)}_{\alpha_0,\alpha_1}(i_1)
	G^{(2)}_{\alpha_1,\alpha_2}(i_2)\cdots
	G^{(N)}_{\alpha_{N-1},\alpha_N}(i_N)
\end{equation}
where $G^{(k)}\in\mathbb{C}^{r_{k-1} \times n_k \times r_k}$ are tensors of order 3, called cores of the TT decomposition. The dimensions $r_k$ are called compression ranks. Since the product of Equation (41) must be a scalar, the constraint $\alpha_0=\alpha_N=1$ must be imposed.  

Turning now to the representation of the thermal wavefunction of Eq. \eqref{eq:tseq} in the TT format, we let $d$ be the number of degrees of freedom of the Hamiltonian operator $H$, and assume that the environment is described using $M$ uncorrelated vibrations.
Hence, when we take $n_\mathrm{b}$ states for each vibration, the vector $\ket{\psi_\theta(t)}$ of Eq. \eqref{eq:tseq} can be considered as a tensor with $N=d+Mn_\mathrm{b}$ indices. Therefore, in Eq. \eqref{eq:ttformat} the first $d$ indices label the DoFs of the system, and the remaining $Mn_\mathrm{b}$ indices label the bath operators.  In Section \ref{sec:result1}, we solve Eq. \eqref{eq:tseq} using the time-dependent variational principle (TDVP) algorithm \cite{LubichEtAl2015SJNA} with a compression rank of 70, which is sufficient for convergence. In Section \ref{sec:result2}, we employ the tAMEn algorithm \cite{Dolgov2019CMAM,BorrelliDolgov2021JPCB,TakahashiBorrelli2024JCTC} with a threshold of $\epsilon = 1.0 \times 10^{-3}$, also ensuring convergence.

\providecommand{\latin}[1]{#1}
\makeatletter
\providecommand{\doi}
  {\begingroup\let\do\@makeother\dospecials
  \catcode`\{=1 \catcode`\}=2 \doi@aux}
\providecommand{\doi@aux}[1]{\endgroup\texttt{#1}}
\makeatother
\providecommand*\mcitethebibliography{\thebibliography}
\csname @ifundefined\endcsname{endmcitethebibliography}  {\let\endmcitethebibliography\endthebibliography}{}


\end{document}